\documentclass[aps,prx,onecolumn,showpacs,superscriptaddress,floatfix,notitlepage]{revtex4-1}  

\usepackage{ucs}
\usepackage{natbib}
\usepackage{graphicx}  
\usepackage{bm}        
\usepackage{amssymb}   
\usepackage{amsmath}
\usepackage{color}
\usepackage{CJK}
\usepackage{times}
\usepackage{verbatim}
\usepackage{mathrsfs}
\usepackage{hyperref}
\usepackage{ulem}
\hypersetup{colorlinks = True, urlcolor=blue, linkcolor=blue, citecolor=blue}
\newcommand{\etal}{\textit{et al}.}

\usepackage{cancel}
\usepackage{comment}
\begin{document}

\title{Disorder upon disorder: localization effects in the Kitaev spin liquid}
\author{Wen-Han Kao}
\affiliation{School of Physics and Astronomy, University of Minnesota, Minneapolis, MN 55455, USA}
\author{Natalia B. Perkins} 
\email{nperkins@umn.edu}
\affiliation{School of Physics and Astronomy, University of Minnesota, Minneapolis, MN 55455, USA}

\date{\today}
\begin{abstract}
In recent years, several magnetic Mott insulators with strong spin-orbit couplings were suggested to be proximate to the Kitaev quantum spin liquid (QSL) whose one of the most exciting features is the fractionalization of spin excitations into itinerant Majorana fermions and static $Z_2$ fluxes. Motivated by the emergence of this plethora of 4d and 5d transition metal Kitaev materials and by the fact that some level of disorder  is inevitable in real materials, here we study how  the Kitaev QSL responds to various forms of disorder, such as vacancies, impurities, and bond randomness. 
First, we argue that the presence of the quenched disorder in the Kitaev QSL can lead to the Anderson localization of Majorana fermions and the appearance of Lifshitz tails.  We point out that the Anderson localization of low-energy states is particularly strong in the extended Kitaev model  with the time reversal symmetry breaking term. Second, we show that the disorder effects on the low-energy Majorana fermion modes can be detected  in thermal transport.
 Third, we show that at finite temperatures the $Z_2$ fluxes become thermally excited and give rise to an additional disorder for the Majorana fermions. This source of the disorder dominates at high temperatures.  Fourth, we demonstrate that both the structure of the energy spectrum and  the thermal transport properties of the  disordered Kitaev QSL depend strongly on the character of disorder.  While we find that both  the site disorder and the bond randomness suppress the longitudinal thermal conductivity, the low-energy localization is stronger in the case of the site disorder.

\end{abstract}
\maketitle

\section{Introduction}\label{intro}

Quantum spin liquid (QSL) is one of the most intriguing states of matter, where interacting spins form a quantum disordered state without spontaneous symmetry breaking, and thus no magnetic long-range order appears even at zero temperature. Its history began in 1973 with  the  pioneering work of  P.W. Anderson \cite{anderson1973resonating}, in which  he proposed that a state consisting of a quantum superposition of spin-singlet states, dubbed Resonating Valence Bond (RVB) state,  might describe the  ground state  of the Heisenberg antiferromagnet on the triangular lattice. While this idea was  proven wrong  as the antiferromagnet has the long-range magnetic order on the triangular lattice \cite{Huse88}, it brought forth a new idea of a state  that can not be written as a product state.  Among  such states the most notable are fractional quantum Hall state in two-dimensional electron gases \cite{Laughlin1983} and QSLs  in magnetic insulators \cite{Kalmeyer1987}. Nowadays the study  of QSLs represents one of the central  problems of interest  in the field of strongly correlated electrons  and several excellent reviews  on QSLs are available \cite{Lee2008,Balents2010,Savary2016, Zhou2017,KnolleMoessner2019,Broholm2020,Takagi2019}. 

 What brought a particular interest to QSLs is 
 a remarkable set of  their emergent  phenomena  including
long-range  entanglement,  topological  ground-state  degeneracy,  and  fractionalized excitations
which can be realized  in them \cite{Kitaev2006,Balents2010,Savary2016, Zhou2017,KnolleMoessner2019,Broholm2020}.  
   Motivated by these intriguing  properties of QSLs, much  work has  been done in identifying candidate materials for realizing  QSLs 
 in real systems.
Recent years have seen much progress in identifying  QSLs  features  on materials where the magnetic ions reside on lattices that frustrate classical magnetic order or when the interactions between them are intrinsically frustrated.
Prominent examples  of QSLs include herbertsmithite and other systems with spin-1/2 copper  ions  occupying the kagome lattice \cite{Norman2016},  a great variety of organic molecular crystals residing on a distorted triangular  lattices \cite{Shimizu2003,Itou2008,Powell2011,Isono2014,Yamashita2017},  the three-dimensional hyperkagome material Na$_4$Ir$_3$O$_8$ \cite{Okamoto2007}, 
honeycomb lattices of Ir  or Ru ions \cite{Jackeli2009,Chaloupka2010,Singh2010,Singh2012,Plumb2014,Sears2015,Rau2016,Trebst2017,Kitagawa2018,Motome2019,Takagi2019}, that are material candidates  for the Kitaev QSL \cite{Kitaev2006}, and many others.

Direct  experimental  observation  and  characterization  of QSLs is challenging. Unlike states with spontaneously broken symmetry, the topological order characteristic of QSLs cannot be captured by a local order parameter and thus cannot be directly detected by local measurements. Identifying QSLs thus relies mainly on the characterization of the excitations of QSL candidates by various  dynamical probes such as inelastic neutron scattering ~\cite{Knolle2015,Knolle2014a,Banerjee2016,Banerjee2017}, Raman scattering \cite{Ko2010,Sandilands2015,Knolle2014b,Nasu2016,Rousochatzakis2019,Sahasrabudhe2020,Yiping2020,Dirk2020} or  resonant inelastic X-ray scattering ~\cite{Gabor2016,Gabor2017,Gabor2019}, making
the phenomenon of fractionalization of  elementary spin excitations into fermionic or bosonic spinons as well as emergent gauge excitations
a   defining  feature of QSLs. 
Thermodynamic and transport measurements of QSLs provide an additional information about the density of states (DOS) and the mobility of the excitations \cite{Katsura2010,Nasu2014,Nasu2015,Hirobe2017,Yamashita2017,Kasaharaprl2018,kasahara2018majorana,Nasu2017,Metavitsiadis2017,Pidatella2019,Motome2020,Widmann2019,LiChen2020,Feng2020}. In particular, all fractionalized quasiparticles contribute into the specific heat reflecting the  total DOS of the system, but only mobile excitations participate in the thermal transport. Therefore in QSLs, in which fractionalization happens into different  kinds of fractionalized quasiparticles, like itinerant Majorana fermions and localized gauge fluxes in the case of the Kitaev spin liquid \cite{Kitaev2006}, the Einstein relation stating that the thermal conductivity of a material is proportional to the thermal diffusion constant and to the specific heat can  be violated.

 One of the long-standing open problems which recently attracted a lot of attention \cite{Willans2010,Willans2011,Watanabe2014,Gabor2014,Zschocke2015, Sreejith2016,savary2017disorder,Yamaguchi2017,kimchi2018heat,Kitagawa2018,Slagle2018theory,li2018role,Knolle2019,Takahashi2019,Do2020, Murayama2020,Yamada2020,Motome2020,Kao2020}
is understanding how QSLs respond to various forms of disorder, such as dislocations, vacancies, impurities, and bond
randomness, which are inevitable in real materials.  It has been noted that
 quenched disorder on top of the quantum disordered strongly correlated spin state of a QSL can give rise to diverse and often puzzling behaviors \cite{Yamaguchi2017,Kitagawa2018,Takahashi2019,Do2020,Murayama2020}. 
 Moreover, given that the properties of QSLs are difficult to detect directly,
  much additional information can be obtained by studying the distinctive
responses to local perturbations, such as static defects, dislocations, and magnetic or non-magnetic impurities. In particular, these perturbations in real materials  may nucleate  excitations characteristic to the QSL under consideration~\cite{Willans2010,Willans2011}. 
Of specific interest is the role of disorder in the materials that have been suggested to be   potential candidates \cite{Rau2016,Trebst2017,hermanns2018,Takagi2019,Motome2019}  to realize the Kitaev QSL~\cite{Kitaev2006}.
In  a flurry of  recent experiments on the honeycomb ruthenium chloride $\alpha$-RuCl$_3$, it was  shown that both   bond disorder and  stacking disorder are not negligible \cite{Plumb2014,Majumder2015,Johnson2015,Sears2015,Banerjee2016}.
Perhaps, disorder  also plays a crucial role for a potential proximity of Ag$_3$LiIr$_2$O$_6$  to a Kitaev QSL state \cite{Tafti2019}. However,  arguably
the most remarkable and intriguing consequences of disorder have been observed in a presumptive quantum spin liquid state of  the hydrogen intercalated iridate H$_3$LiIr$_2$O$_6$
 \cite{Kitagawa2018}.

  Much of the intuition on the effect of disorder on the low-energy properties of QSL can be obtained by  analogies with 
the disorder effects on the single-particle electron wavefunctions in solids, the study of which,  not surprisingly, was also  pioneered by 
P.W. Anderson back in 1958 \cite{Anderson1958}. He showed that  the wavefunction of non-interacting quantum particles on the lattice may be exponentially localized near  some point in space due to a random potential, provided that the randomness is sufficiently strong. In the early 1960s Nevill Mott  described the transition between the delocalized and localized states with the help of the notion of a mobility edge \cite{Mott1967}. 
The later  studies have shown that in non-interacting one- (1D)  and two-dimensional (2D) systems even weak
disorder localizes all electronic states \cite{Abrahams1979}, thus leading to the exactly zero conductivity.  The current understanding is that a true phase transition between itinerant and localized states, known as the Anderson transition, can exist only in three dimensions and that it requires rather strong disorder \cite{Abrahams1979}. The localization problem becomes more difficult if one goes beyond the picture of non-interacting particles since interactions, and in particular repulsive interactions between electrons, can destroy localization and lead to more complex phenomena.



The phenomena of  Anderson localization and  Anderson transition were intensively studied and applied to various systems, see, e.g. excellent books and reviews \cite{Lee1985,Boris1984,Kramer1993,50Anderson,Evers2008}.  
A comprehensive description of the experimental and theoretical   developments of Anderson localization and Anderson transition ideas  during the first 50 years after the original work
can be found in the book edited by  Abrahams \cite{50Anderson}, in which a group of experts contributed their personal insights on the subject.

The question which we address in this paper is whether in the presence of quenched disorder we can have  a phenomenon similar to the Anderson localization  in a  QSL. For concreteness and for the simplicity  of the analysis we will focus on the impact of disorder on the properties of the Kitaev QSL, which  is realized in  a  system  of  spin-1/2  at  sites  of a honeycomb lattice interacting via Ising-like frustrated nearest-neighbor exchange interactions \cite{Kitaev2006}. This model   is exactly solvable, has  a QSL ground state, and  is yet realistic \cite{Kitaev2006,Baskaran2007,Jackeli2009}.
Spin excitations in the Kitaev model are  fractionalized  into  two  very different types  of  quasiparticles: itinerant spinon-like excitations, which are described by the Majorana fermions which are gapless or gapped depending on the coupling parameters, and localized gapped $Z_2$ gauge fluxes \cite{Kitaev2006}. Because  the $Z_2$ gauge fluxes do not have any dynamics, 
different flux sectors can be considered independently, which is a great simplification of the problem and the essence of its exact solvability.
In each  of the flux sectors, the model effectively reduces to the 
 free-fermion Hamiltonian describing the hopping of  the itinerant  Majorana fermions.  The fact that the  Majorana fermions are non-interacting and that this remains true even in the presence of   various types of quenched disorder, 
 makes the Kitaev QSL  an ideal setting for exploring   novel disorder-induced  localization effects on a quantitative level.

In this work, we  consider  three types of  disorder in the Kitaev model: the  bond  disorder, the site disorder  and the thermal disorder \cite{Knolle2019,Yamada2020,Motome2020,Kao2020}. In real materials, the bond disorder can arise from random lattice
distortions and/or chemical disorder on non-magnetic sites, both of which locally modify
individual exchange paths.  The site disorder
 can also originate from various sources, such as missing magnetic moments or the
presence of non-magnetic impurities (true vacancies), or from local weak couplings of
magnetic moments due to strong but rare bond randomness (quasivacancies). 
In our previous works we showed 
that introducing bond and site disorder in the Kitaev QSL preserves
most of the spin-liquid behavior but leads to distinct changes in the low-energy physics \cite{Knolle2019,Kao2020}. 
Crucially, two types of disorder affect two types of the fractionalized excitations very differently.
Bond disorder  leads both to the reduction of the flux gap \cite{Zschocke2015, Yamada2020} and to a pileup of low-energy modes which cause a distinctive power-law divergence in the fermionic density of states \cite{Knolle2019}. 
The effect of the site disorder  is also well 
pronounced  even at very low concentration of vacancies or quasivacancies,  i.e., with weak disorder.
Vacancy-induced Majorana modes are accumulated in a low-energy peak of the DOS across a broad window at low energies, which  is well fitted by some power-law with an exponent 
  determined by the nature and concentration of the vacancies. Moreover, the
presence of site disorder leads to the partial localization of  itinerant Majorana fermions near the vacancy centers  \cite{Kao2020}.  However in both cases, the “pure” Dirac dispersion of $Z_2$ Dirac spin liquid is lost. The third type of disorder, the thermal disorder,
  is a distinguished thermodynamic property  of  the  Kitaev  model.
  At finite temperatures, the fluxes become thermally excited and give rise to an additional disorder for the itinerant Majorana fermions. The separated energy scales of flux excitations and itinerant Majorana fermion excitations lead to two finite-temperature crossovers in the specific heat, known as the thermal fractionalization of spins in the Kitaev honeycomb model \cite{Nasu2015}. At high temperatures, the disorder from thermal fluxes flattens the Majorana density of states over the whole energy range, and its effect overcomes  those from the quenched disorders. 
  

 The rest of the paper is organized as follows:  In Sec.\ref{DisorderKSL}, we  begin  with a  discussion of 
 fractionalization  in the Kitaev honeycomb model, and then introduce different types of quenched disorder which we consider in this work.
We then proceed in Sec.\ref{DOS} to analyze in detail how the presence of various types of disorder affects the low-temperature density of  Majorana fermion states. For each scenario of quenched disorder we compute  the inverse participation ratio (IPR), which allows us to capture the localized nature of the low-energy eigenstates and the appearance of Lifshitz tails at the high-energy edge of the Majorana fermion band. Since the localization properties of the states clearly influence the transport
properties of the system and since the Kitael QSL is an insulator, in Sec.\ref{thermal-conductivity} we study how they reveal themselves   in the thermal transport. In particular, we compute
the longitudinal thermal conductivity and show that different type of disorder have distinct effect on the
 Drude weight,   which measures the non-dissipative contribution to the heat flow, and  on the temperature-dependent  thermal conductivity 
  coefficient, which can be obtained  from the zero-frequency extrapolation  of the dynamical part of the thermal conductivity. 
Finally, in Sec.\ref{Conclusion}, we summarize the main results of this paper.

\begin{figure*}
     \includegraphics[width=1.0\textwidth]{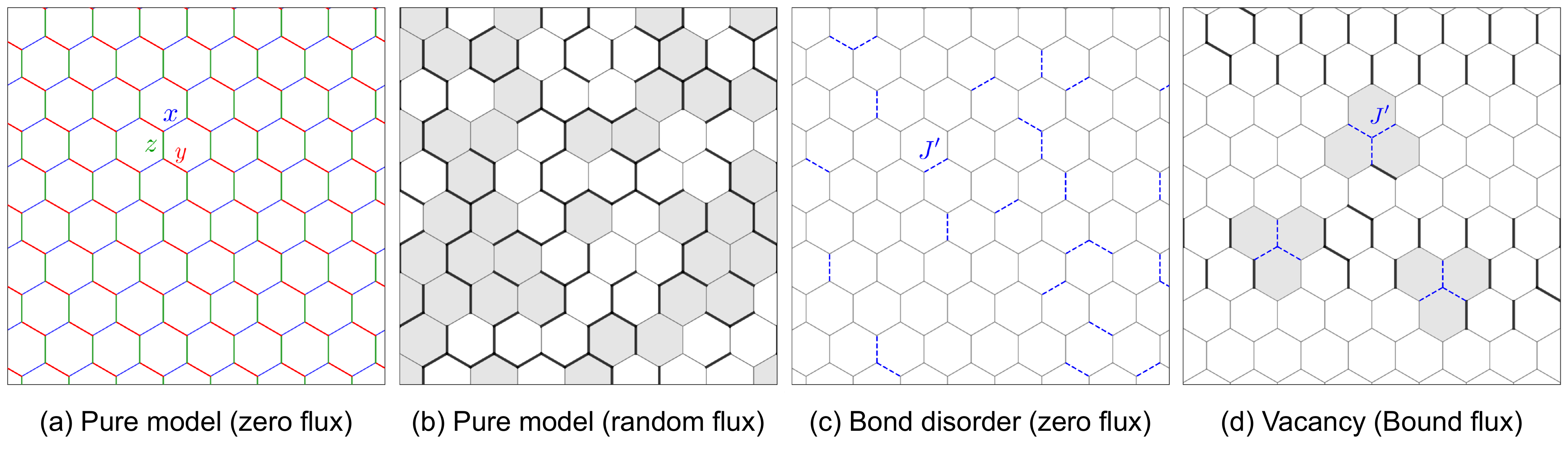}
     \caption{\label{fig:lattice_graphs} Disordered Kitaev honeycomb model. (a) Pure model with zero-flux sector. In this work the isotropic coupling is considered: $J_{x} = J_{y} = J_{z} = J$. (b) Pure model with random-flux sector. The black thick lines represent $u = -1$. The flux on a hexagon ($W_{p} = -1$) is depicted by shaded hexagon. (c) Bond disorder with zero-flux sector. Couplings with density $\rho_{b}$ are replaced by $J^{\prime} \neq J$, where $J^{\prime}$ can be a constant value or a random number obeying specific distributions. (d) Vacancy with bound-flux sector. True vacancies refers to $J^{\prime} = 0$ and quasivacancies refers to $J^{\prime} \ll J$.}
\end{figure*}

\section{Disordered Kitaev model}\label{DisorderKSL}

We focus our discussion on a minimal exactly soluble model ~\cite{Kitaev2006}
\begin{align}
\label{eq:Hamspin}
  \mathcal{H} = -\sum_{\left \langle ij \right \rangle}J_{\left \langle ij \right \rangle_{\alpha}}\hat{\sigma}_{i}^{\alpha}\hat{\sigma}_{j}^{\alpha} - h\sum_{\left \langle \left \langle ik \right \rangle \right \rangle}\hat{\sigma}_{i}^{\alpha}\hat{\sigma}_{j}^{\beta}\hat{\sigma}_{k}^{\gamma},
\end{align}
where $\hat{\sigma}^{\alpha}_{i}$ denotes Pauli spin operators with $\alpha = x, y, z$ and $\left \langle ij \right \rangle_{\alpha}$ labels the nearest-neighbor sites $i$ and $j$ along an $\alpha$-type bond.   
In the pristine Kitaev model, we only consider the isotropic coupling with $J_{x} = J_{y} = J_{z} = J$ [see Fig.\ref{fig:lattice_graphs}(a)].
The second term is the  three-spin interaction with strength $h \sim \frac{H_{x}H_{y}H_{z}}{J^{2}}$, which is the leading-order perturbative contribution from the Zeeman field $H$. This term imitates the external field effect and breaks  time-reversal symmetry while preserving the exact solution of the model \cite{Kitaev2006}. By rewriting each spin operator in terms of four Majorana fermions, $\hat{\sigma}^{\alpha}_{i} = i\hat{b}^{\alpha}_{i}\hat{c}_{i}$, and defining the link operators $\hat{u}_{ij}=i\hat{b}^{\alpha}_{i}\hat{b}^{\alpha}_{j}$, the Hamiltonian takes the form
\begin{align} \label{eq:HamMF}
 \mathcal{H} = i\sum_{\left \langle ij \right \rangle}J_{\left \langle ij \right \rangle_{\alpha}}\hat{u}_{\left \langle ij \right \rangle_{\alpha}}\hat{c}_{i}\hat{c}_{j} + ih\sum_{\left \langle \left \langle ik \right \rangle \right \rangle}\hat{u}_{\left \langle ij \right \rangle_{\alpha}}\hat{u}_{\left \langle kj \right \rangle_{\beta}}\hat{c}_{i}\hat{c}_{k}.
\end{align}

First, to account for a  bond disorder, we randomly place nearest-neighbor couplings with different strength $J^{\prime}$ in the pristine system with $J_{\langle ij \rangle_\alpha} = J = 1$. Two tuning parameters are given: the distribution of random couplings $J^{\prime}$ and the doping concentration $\rho_{b}$ (see Fig.\ref{fig:lattice_graphs}(c) for one of the bond-disordered realizations). In the simplest case of \textit{constant bond disorder}, the system is doped with weaker or stronger bonds with constant strength $J^{\prime} \neq 1$. In \textit{uniform bond disorder}, the random coupling $J^{\prime}$ is governed by the uniform (box) distribution $J^{\prime} \in [J-\delta J, J+\delta J]$ such that any coupling strengths within this interval are equally populated. Lastly, in the \textit{binary bond disorder}, the random bonds can be either stronger or weaker than $J$ by a constant value $\delta J$, namely $J^{\prime} = J \pm \delta J$.

Second, to introduce randomly distributed vacancies into the Kitaev honeycomb model,  we rewrite the first term in (\ref{eq:HamMF}) as 
\begin{equation}\label{eq:HamMFvac}
    \mathcal{H} = i\sum_{\substack{\left \langle ij \right \rangle\\ i,j \in \mathbb{P}}}J_{\left \langle ij \right \rangle_{\alpha}}\hat{u}_{\left \langle ij \right \rangle_{\alpha}}\hat{c}_{i}\hat{c}_{j} + i\sum_{\substack{\left \langle kl \right \rangle\\ k \in \mathbb{V}, l \in \mathbb{P}}}J^{\prime}_{\left \langle kl \right \rangle_{\alpha}}\hat{u}_{\left \langle kl \right \rangle_{\alpha}}\hat{c}_{k}\hat{c}_{l},
\end{equation}
where $\mathbb{P}$ denotes the subset of normal lattice sites and $\mathbb{V}$ denotes the subset of vacancy sites. We consider a 
compensated case  with equal   numbers of vacancies on the two sublattices of the honeycomb lattice.
By taking the limit of $J^{\prime}_{\alpha} \ll J_{\alpha}$, sites belonging to $\mathbb{V}$ behave as quasivacancies (see Fig.\ref{fig:lattice_graphs}(d) for one of the configurations with random  quasivacancies). In the limit of $J^{\prime}_{\alpha}\rightarrow 0$, quasivacancies become true vacancies in which
 a Majorana fermion $\hat{c}$ remains on the vacancy site, but its nearest-neighbor hopping amplitudes are removed.
 
 The solvability of the Kitaev model relies on the extensive number of conserved fluxes defined on each hexagonal plaquette,
$\hat{W}_{p} = \hat{\sigma}^{x}_{1}\hat{\sigma}^{y}_{2}\hat{\sigma}^{z}_{3}\hat{\sigma}^{x}_{4}\hat{\sigma}^{y}_{5}\hat{\sigma}^{z}_{6} 
    = \prod_{\left \langle ij \right \rangle \in p}\hat{u}_{\left \langle ij \right \rangle_{\alpha}}$,
which can block-diagonalize the Hamiltonian (\ref{eq:Hamspin}) into flux sectors since fluxes commute with each other, $[\hat{W}_{p}, \hat{W}_{p'}] = 0$, and with the Hamiltonian, $[\hat{W}_{p}, \mathcal{H}] = 0$. Both the flux operators $\hat{W}_{p}$ and the link operators $\hat{u}_{\left \langle ij \right \rangle_{\alpha}}$ have eigenvalues $\pm 1$.  Note that not all choices of 
$\{
\hat{u}_{\left\langle ij \right \rangle_\alpha} \}$ correspond to distinct physical  states  of  the  spin  model,  and  only  those that  are  gauge  inequivalent  should  be  treated  as distinct.  

Once the link variable is specified for each bond (see Fig.\ref{fig:lattice_graphs}(b) for a random flux configuration in the pure Kitaev model) and the physically relevant flux sector is determined, the Hamiltonian  (\ref{eq:HamMF}) can be  solved exactly as a tight-binding model of Majorana fermions. This remains true in the presence of disorder, even though the number of flux degrees of freedom in the presence of true vacancies is effectively reduced \cite{Kao2020}.
  In  all our calculations, the diagonalization of the Hamiltonian  (\ref{eq:HamMF}) is performed numerically  on the finite size cluster with periodic boundary conditions. The resulting diagonal form  is given by
 \begin{align}\label{eq:HamMF-diag}
\mathcal{H} = \sum_{n}\epsilon_{n}(\hat{a}_{n}^{\dagger}\hat{a}_{n}-\frac{1}{2}),
\end{align}
where $\hat{a}_n$ are complex matter fermions (superposition of two Majorana operators) which label the eigenmodes  with the fermion energies $\epsilon_n \equiv\epsilon_n \left (\{{ J_{{\langle ij \rangle}_{\alpha}}}\},\{\hat{u}_{\left \langle ij \right \rangle_{\alpha}}\}\right)$ for a given realization of disorder in a given flux sector.  
 Also, the energy of the  lowest-energy state in a given flux configuration for a particular realization of disorder,
 $E^{(0)}_{f}\equiv -\frac{1}{2} \sum_{n}\epsilon_{n}$,  which corresponds to all unoccupied fermionic eigenmodes, is associated with the energy of  a corresponding flux sector.
We recall that at finite temperatures, the model (\ref{eq:HamMF}) can be regarded as a model  of  noninteracting Majorana fermions coupled to thermally excited $Z_2$ fluxes, and therefore the thermally acivated disorder is present even in the pristine model \cite{Nasu2014,Nasu2015,Motome2019,Feng2020}.

\section{Density of states: features of localization}  \label{DOS}

 In this section, we discuss how the presence of disorder affects the low-temperature density of  Majorana fermion states. We consider the following  cases: (a) true vacancies with $J' = 0$, (b) quasivacancies with $J' > 0$, (c) constant bond disorder with weaker (stronger) bonds $J' = 0.5 \,(2.0)$ and $\rho_{b} = 10\%$, (d) uniform bond disorder with $\delta J = 0.3, 0.5, 0.8$ and $\rho_{b} = 100\%$, and (e) binary disorder with $\delta J = 0.8$ and $\rho_{b} = 25\%$, which is considered in the previous paper by Knolle \etal ~\cite{Knolle2019}. Note that in (a) and (b) we consider the bound-flux sector. 
 The algorithm we implement for creating bound-flux sectors  (see Fig.\ref{fig:lattice_graphs} (d) for an illustration of the bound-flux sector) follows the recipe in Ref.~\onlinecite{Kao2020}. 
  Previous studies \cite{Willans2010,Willans2011, Kao2020}
  show that in the  Kitaev system  with the presence of  small density of vacancies, the total energy is lower when fluxes are bound to the vacancies than in the zero-flux case, and 
  that the flux-binding effect remains  in the case of quasivacancies for $J^{\prime}/J < 0.0544$ \cite{Kao2020}.
   In (c) and (d), we apply zero-flux sector in the calculation, since in the presence of weak disorder, the low-energy sector is still the zero-flux sector. However, for strong disorder case, some of the local flux gaps tend to decrease or even vanish \cite{Zschocke2015, Yamada2020}, thus it might be more reasonable to compare the two extreme cases: zero-flux sector and random-flux sector. In case (e), we will present the results for both sectors.

\subsubsection{Density of states and inverse participation ratio}

\begin{figure*}
     \includegraphics[width=1.0\textwidth]{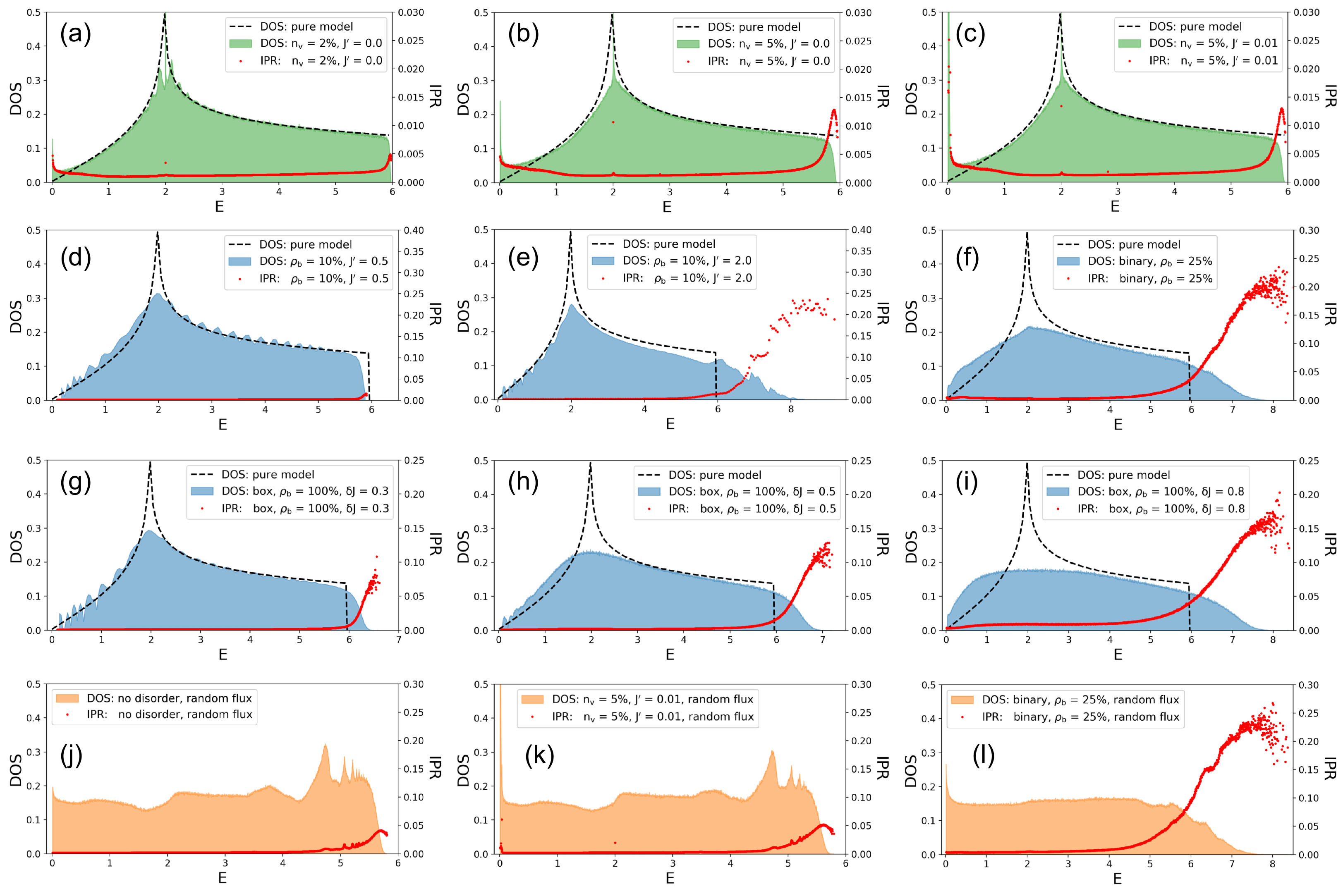}
     \caption{\label{fig:DOS_IPR} Density of states (DOS) and inverse participartion ratio (IPR) for various types of quenched disorder. (a)-(b): True vacancy with $J^{\prime} = 0.0$ in the bound-flux sector. (c): Quasivacancy with $J^{\prime} = 0.01$ in the bound-flux sector. (d)-(e): 10\% random bonds with weaker or stronger coupling constant, also denoted as $J^{\prime}$. Zero-flux sector is considered. (f): 25\% random bonds with binary disorder of $\delta J = 0.8$ in the zero-flux sector. (g)-(i): 100\% random bonds with box distribution of different widths. Zero-flux sector is considered. (j)-(l): Comparison among pure model, system with quasivacancies or binary bond disorder in random-flux sectors. All results are calculated from one $L = 40$ cluster and averaged over 2000 random realizations.}
\end{figure*}

For each scenario of quenched disorder, the density of states (DOS) is given by 
\begin{equation}\label{Eq:DOS_def}
    N(E) =  \left \langle \sum_{n}\delta(E-\epsilon_{n})\right \rangle_{\left \{ \left\{ J_{\langle ij \rangle_{\alpha}}\right \},\, \left \{ \hat{u}_{\left \langle ij \right \rangle_{\alpha}}\right \} \right \}} ,
\end{equation}
where the brackets refer to the average over independent disordered samples and the average over flux sectors (if different flux sectors are considered).  Different types of disorder modify the DOS in a different way [compare various panels of Fig.~\ref{fig:DOS_IPR}], however there is one common trend: the  van Hove singularity of the pristine model is destroyed by any type  of disorder, and especially by the thermal disorder. 

The DOS  by itself, however, can not reflect how strong the localization effect is for each state. Thus we need to introduce another quantity to describe the localization phenomena.
The localized nature of the low-energy eigenmodes can be illustrated by the inverse participation ratio (IPR). This quantity is defined as
\begin{equation}\label{Eq:IPR_def}
    \mathcal{P}_{n} = \sum_{i}|\phi_{n,i}|^{4},
\end{equation}
where the index $n$ labels the eigenmode  wave function $\phi_{n,i}$ and the index $i$ labels the lattice site. For a delocalized mode, the IPR scales roughly as $\sim 1/N$ in a system with $N$ sites since the wavefunction is spread out uniformly over the entire lattice. This behavior is precisely what we see  for the fermionic bulk modes \cite{Kao2020}. However, for the low-energy modes realized in the presence of some forms of disorder, the IPR is significantly larger since the wave function is confined to a small portion of the lattice \cite{Kao2020}. This quantity was also used in studies of disordered graphene as an indicator of vacancy-induced quasilocalized modes \cite{Pereira2008}. As we will discuss later in more details, the IPR is also size-dependent \cite{Kramer1993}, and the question remains whether or not it remains finite in the thermodynamic limit.


\subsubsection{Site disorder: vacancies and quasivacancies}\label{Site-disorder}

 In Fig.~\ref{fig:DOS_IPR} (a)-(c), 
the presence of low concentration of vacancies introduces a pileup of low-energy Majorana fermion states in the  bound-flux sector, but the rest of the DOS remains similar to the pure model in the zero-flux sector. Thus, this type of disorder is a weak disorder. 
The IPR shows that the low-energy states have higher level of localization compared to the bulk modes. The amplitude plot of 
the real-space wavefunctions of the low-energy modes introduced by the vacancies presented in Fig.~\ref{fig:mode_amplitudes} (a) further supports the IPR analysis.  It shows that
the amplitude of these lowest energy wave functions
is slightly larger around the vacancies, but still it spreads out roughly as $1/r^{\alpha}$ where $\alpha$ is smaller than 1. Thus, these low-energy modes are only  quasilocalized. In Ref.~\onlinecite{Kao2020}, we have argued that these quasilocalized states give rise to visible effects in the thermodynamic quantities such as the specific heat. 
The IPR also shows that we observe the famous Lifshitz tails~\cite{Lifshitz1965} for the Majorana fermion states near the top band edge.   These states are localized within finite region of the lattice (quasilocalized) and thus have finite IPRs.   The states  in the middle of the Majorana's band are  delocalized  and have  vanishingly small IPRs.

\subsubsection{Various types of bond disorder}  
\label{Bond-disorder}
\begin{figure*}
     \includegraphics[width=1.0\textwidth]{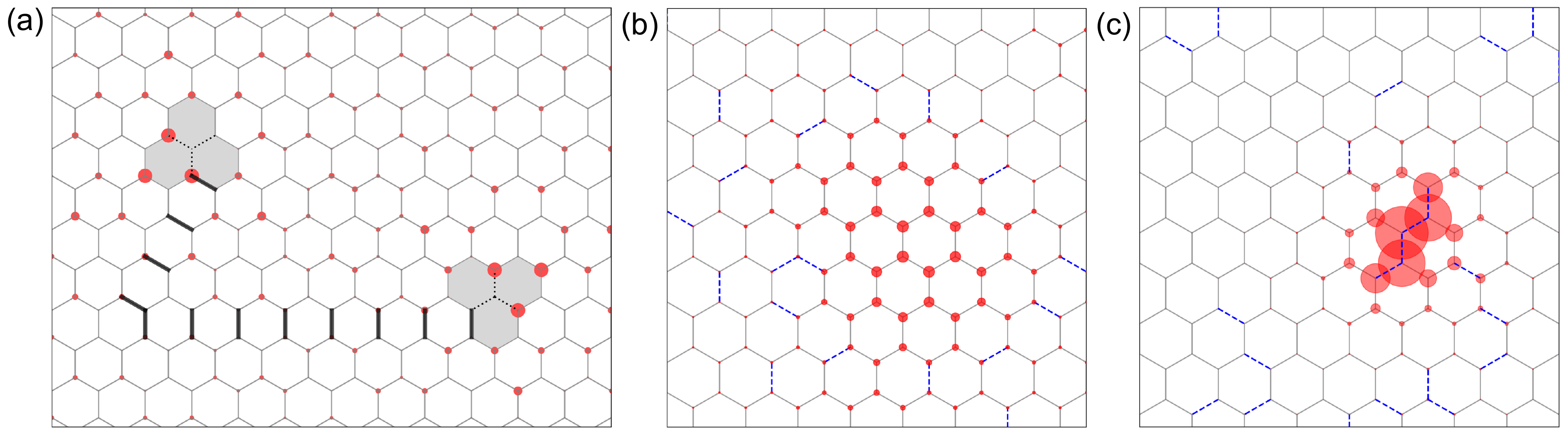}
     \caption{\label{fig:mode_amplitudes} Quasilocalized modes in the presence of quenched disorder. (a) Low-energy eigenmode with vacancies and bound-flux sector. (b) High-energy eigenmode with weaker random bonds, corresponding to Fig.~\ref{fig:DOS_IPR} (d). (c) High-energy eigenmodes with stronger random bonds, corresponding to Fig.~\ref{fig:DOS_IPR} (e). The dotted lines depict the removed bonds in the case of vacancies, and the dashed lines depict the bonds with weaker or stronger strength in the case of bond disorder. The real-space amplitudes are shown as red filled circles. The black thick lines represent the flipped link $u = -1$, which is the eigenvalue of local operator $\hat{u}$.}
\end{figure*}

In Fig.~\ref{fig:DOS_IPR} (d)-(e), we consider a case of the bond-disorder in the zero-flux sector. To this end, we  replace 10\% of the nearest-neighbor couplings of the strength $J$ in the pristine model with weaker or stronger couplings $J'$. In the weak-bond case [Fig.~\ref{fig:DOS_IPR} (d)], we see that the DOS increases slightly in the region below the van Hove singularity at $E = 2$, which is not surprising since we have regions with weaker interaction. The changes in the IPR compared with the IPR in the pristine model \cite{Kao2020} are also very small: 
 only the states near the   band edge show a tiny increment, indicating a weak localization at the high-energy edge. However, in the strong-bond case (see Fig.~\ref{fig:DOS_IPR} (e)), Lifshitz-like tails with strongly localized states appear near the band edge. Similar effect was discussed in the original Anderson model of single-particle localization \cite{Anderson1958}. The dramatic difference in IPR between the weak-bond and strong-bond cases can be understood in terms of the real-space wavefunctions. For the high-energy states near the band edge, the amplitudes are spread out mostly on the strong-coupling region. 
 For example, in the case of the doping with weaker bonds (see Fig.~\ref{fig:mode_amplitudes} (b)), the stronger bonds are those normal majority bonds with $J = 1$. Thus the wavefunction of high-energy states is spread over the normal bonds and confined by the presence of weak doped bonds (blue dashed lines in Fig.~\ref{fig:mode_amplitudes} (b)). Since only 10\% bonds are doped and weakened, these states are still quite delocalized. On the contrary, when the stronger bonds are doped (Fig.~\ref{fig:mode_amplitudes} (c)), the wavefunction of high-energy states in the tail of the band spreads over the doped bonds $J'$, and because of the low-concentration of those bonds, the wavefunction is much more localized and shows much larger IPR in Fig.~\ref{fig:DOS_IPR} (e).


In Fig.~\ref{fig:DOS_IPR} (g)-(i), all the couplings in the system are assigned with random numbers in a uniform (box) distribution in the range $J_{ij}\in [J-\delta J, J+\delta J]$ with $\delta J=0.3, \,0.5$ and 0.8 in (g), (h)  and (i), respectively.   The results for the DOS and the IPR are presented in the zero-flux sector. By increasing the width $2\delta J$ of the distribution, the level of disorder can be enhanced. Similar to the case of doped bonds, the DOS presents clear Lifshitz tail at the high-energy edge, and the IPR of the corresponding states is pretty large. This behavior is consistent with the observation of a recent Monte Carlo study by Nasu \etal  \cite{Motome2020}, which shows that larger width of box distribution leads to the suppression of longitudinal thermal conductivity in the high-temperature region.

In Fig.~\ref{fig:DOS_IPR} (f)  we present the DOS and the IPR results for the case
of the binary disordered bonds, previously considered in Ref. \onlinecite{Knolle2019}, in the zero-flux sectors, respectively. Here  25\% of the bonds are replaced by a different value $J\pm \delta J$, where  $\delta J=0.8$. This  type of disorder 
 corresponds to a strong disorder, which is clearly seen from  rather significant  modification of the DOS  in the whole range of energy eigenstates.  Contrary to the vacancy disorder in the bound-flux sector, no power-law upturn in the DOS is seen at low energies. Moreover,  partial localization of states, indicated by pretty large IPR values for the corresponding states, is only seen at the top of band.
 
 \subsubsection{Thermal disorder  }\label{Thermal-disorder}
In order to clarify the effect of solely thermal disorder, which is the dominant disorder
at high-enough temperatures,
 in Fig.~\ref{fig:DOS_IPR} (j) we  present the results for the DOS and the IPR averaged over random-flux sectors for the pristine Kitaev model. The random fluxes, acting as random couplings of the opposite sign, flatten the overall Majorana fermion's DOS in the whole range of allowed energies. Thermal disorder, however, leads only to weak  localization effects  of the  states near the high-energy band edge. 

\begin{figure*}
     \includegraphics[width=1.0\textwidth]{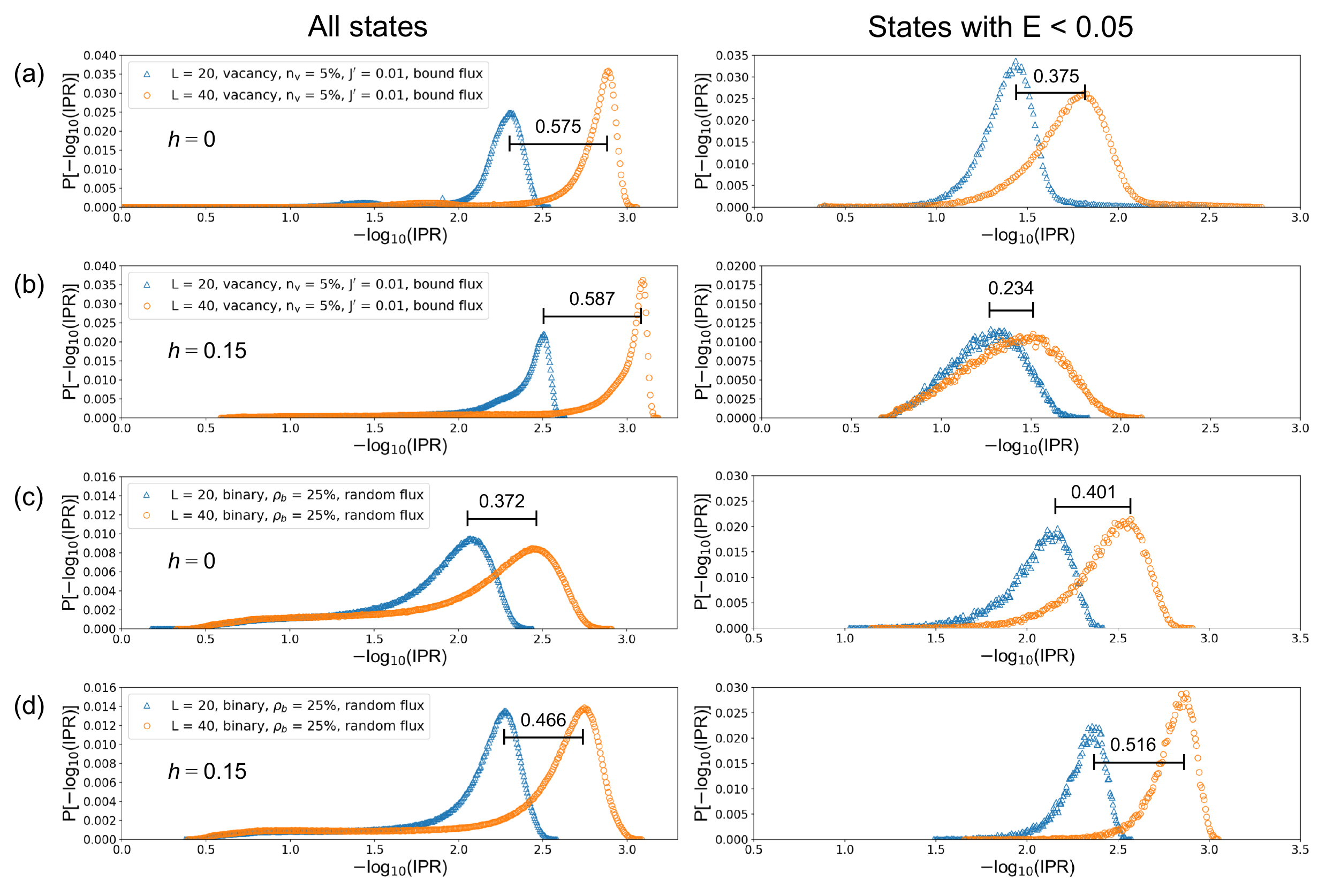}
     \caption{\label{fig:IPR_distribution} Distribution of inverse participation ratio (IPR). The left panel shows the distribution of IPR for all eigenstates, and the right panel is for low-energy states with $E < 0.05$. (a)-(b) 5\% quasivacancies with $J^{\prime} = 0.01$ and bound-flux sector. The presence of three-spin interaction ($h$) makes the low-energy states more localized. (c)-(d) 25\% binary bond disorder with $\delta J = 0.8$ and random-flux sector. The presence of $h$ leads to smaller IPR for all states.}
\end{figure*}

Finally,  we study numerically 
 the interplay  between the site disorder and  the thermal disorder and the binary disorder  and  the thermal disorder. In Fig.~\ref{fig:DOS_IPR} (k) and (l)  we plot the DOS and the IPR averaged, respectively, over different vacancy and  bond disorder configurations and independent  flux sectors.  In both cases, a power-law upturn in  the DOS is seen  at very low-energies \cite{Knolle2019,Kao2020}. However,  in both cases no significant localization effect  is observed  for the low-energy states: the IPR values remain pretty small. Note that the localization of states near the top of the band is seen only for the binary disorder, which is not surprising given the fact that this is a strong disorder in  the Kitaev spin liquid, compared to the weak disordered case with low concentration of vacancies.


\subsubsection{Quenched disorder effects on the low-energy states }\label{low-energy}

Next, we focus on the properties of the low-energy states induced by the quenched disorder. As we discussed above, both the binary bond disorder with random fluxes and the site disorder with bound fluxes can lead to noticeable pileup of low-energy states \cite{Knolle2019, Kao2020}. However, comparing to higher energy states, the relative level of localization and the response to three-spin term (see below) are different in these two cases. With site disorder, the IPR is enhanced by the vacancy-induced states at $E < 0.05$ [see Fig.~\ref{fig:DOS_IPR} (b) and (c)]. On the other hand, 
the IPR for the pileup of states at $E < 0.05$ seems to be small in the case of binary disorder (see Fig.~\ref{fig:DOS_IPR} (l)). 


In Fig.~\ref{fig:IPR_distribution}, we present the distribution of IPR in the log-log scale  calculated for two different system sizes for both cases: panels (a) and (b) for the site disorder and  panels (c) and (d) for  the binary bond disorder.
Generally speaking, if eigenmodes are completely delocalized, the IPR shows $\sim 1/N$ behavior and the peak of IPR population will shift when changing the system size. For example, when systems with $L = 20$ ($N = 800$) and $L = 40$ ($N = 3200$) are considered, $-\log_{10}(\mathrm{IPR})$ of a purely delocalized state should shift to the right by $\log_{10}4 \sim 0.602$. On the left panel of Fig.~\ref{fig:IPR_distribution} (a), we see that the shift of the most populated peak is close to $\log_{10}4$, indicating that most of the states are delocalized. On the right panel, we show the IPR distribution only for states with $E < 0.05$, and the result of $0.375$ indicates that those low-energy states are more localized than the rest. When turning on the three-spin interaction  with the strength $h$ in Eq.(\ref{eq:Hamspin}), a bulk gap in the DOS opens but the vacancy-induced states remain inside the gap. In the bound-flux sector, it was shown that vacancy-induced states appear even around $E \sim 0$ \cite{Kao2020}. As shown on the left panel of Fig.~\ref{fig:IPR_distribution} (b), the IPR distribution
 of states in the bulk shows that these states   become even more delocalized since the second-nearest-neighbor hopping of Majorana fermions leads to the additional delocalization. 
 However, the effect of $h$ on the low-energy states is the opposite -- the shift of the peak positions  $\sim 0.234$ of  the IPR  distributions for two system sizes shown on the right panel of Fig.~\ref{fig:IPR_distribution} (b) indicates stronger localization
 of low-energy states.

With binary bond disorder and random fluxes (see Fig.~\ref{fig:IPR_distribution} (c)), the peak shift ($0.372$) is relatively smaller than in the site disorder case, since the IPR of most eigenmodes is enhanced. However, in the low-energy region ($E < 0.05$), the peak shift becomes larger and the states in the pileup are more delocalized. Furthermore, in the presence of three-spin interaction (see Fig.~\ref{fig:IPR_distribution} (d)), all the states become more delocalized, which is a distinctive feature between the bond disorder and site disorder.

Interestingly, when the antiferromagnetic random couplings $J^{\prime} < 0$ are introduced, a similar effect to the thermal disorder that flatten the overall DOS can be shown numerically. We put this additional discussion in the appendix.

 \section{Thermal conductivity}\label{thermal-conductivity}
  We now turn to the question of whether  the disorder effects on the low-energy Majorana modes can be detected  in thermal transport \cite{Katsura2010,Metavitsiadis2017,Pidatella2019,Nasu2017,Motome2020}.  Here we assume that in  realistic thermal conductivity measurements the lattice contribution can be effectively subtracted,  and  thus we  neglect phonons and assume that the thermal conduction happens  solely through itinerant Majorana   fermions. The explicit  derivation of the thermal conductivity  in the Kitaev QSL 
  was done   by Nasu and Motome \cite{Nasu2017}, and in our work we use the same  formulation.

   The thermal conductivity  is usually computed  in a linear response theory by using the Kubo formula,
  \begin{eqnarray}
  \kappa_{\mu\nu}(\omega, T)=\frac{1}{TV}\int_{0}^\infty e^{i\omega t}\int_{0}^{\beta} d\lambda \langle  J^\nu(-i\lambda)J^\mu(t)\rangle,
  \end{eqnarray} 
  where $\beta=1/T$ is the inverse temperature, 
    $V$ is the volume of the system and 
the energy (heat) current, $J^\nu(t)=e^{i   \mathcal{H}  t} J^\nu e^{-i   \mathcal{H}  t}$,    is defined through the  derivative of the energy polarization operator, ${\bf P}_E=\sum_{j,j'}1/2\,({\bf r}_j+{\bf r}_{j'})  \mathcal{H}_{j,j'}$, as ${\bf J}=d{{\bf P}_E}/dt=i[ \mathcal{H}, {\bf P}_E]$. 
 The explicit expression of the energy current in terms of Majorana fermions is given by
\begin{equation}\label{eq:energy_current}
    \mathbf{J} = i\sum_{\left \langle \left \langle ik \right \rangle \right \rangle}
    J_{\left \langle ij \right \rangle_{\alpha}}J_{\left \langle kj \right \rangle_{\beta}}\hat{u}_{\left \langle ij \right \rangle_{\alpha}}\hat{u}_{\left \langle kj \right \rangle_{\beta}}\left(\frac{\mathbf{r}_{k}-\mathbf{r}_{i}}{2}\right)\hat{c}_{i}\hat{c}_{k},
\end{equation}
  where $\left \langle \left \langle ik \right \rangle \right \rangle$ labels the second-neighbor pair of sites $i$ and $k$ which are connected by the intermediate site $j$. Eq. (\ref{eq:energy_current}) explicitly shows that the energy current  is determined  not only by the states of the itinerant Majorana fermions but also depends on  the localized $Z_2$ variables. Moreover it shows that the current operator involves the hopping of Majorana fermions between second nearest neighbors.
 
\begin{figure*}
     \includegraphics[width=1.0\textwidth]{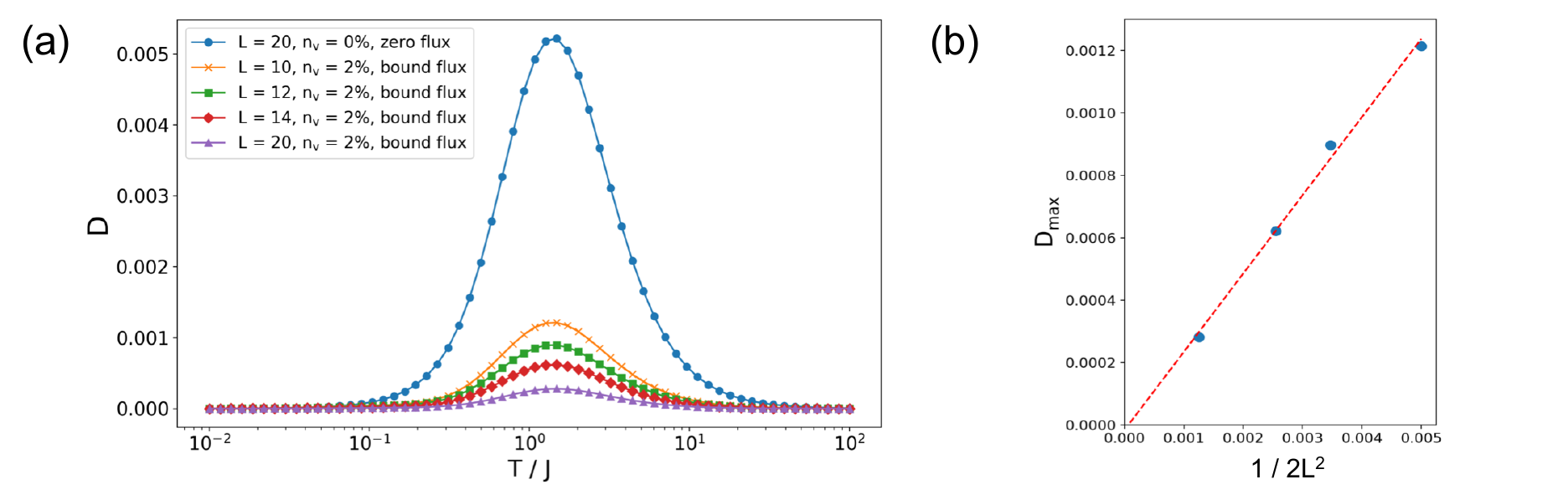}
     \caption{\label{fig:Drude_weight} Drude weight with static flux background. The peak around $T/J \sim 1$ vanishes in the thermodynamic limit. The calculations are done for a superlattice $N = 100$ and cluster size $2L^{2} = 800$, with average over 50 disorder samples.}
\end{figure*}

  We start with the longitudinal component of the
thermal conductivity, $\kappa_{xx}(\omega, T)$, in the time-reversal symmetric Kitaev model ($h=0$).
  Following the work by Nasu and Motome \cite{Nasu2017,Motome2020}, we numerically compute  $\kappa_{xx}$ by introducing a superlattice of $N$ unit cells consisting of $2L^2$-site clusters, such that the site index $i$ of each Majorana fermion decomposes into unit-cell index $l$ and site index $s$ within a unit cell. Then the Hamiltonian of the system can be written as
\begin{equation}
    \mathcal{H} = \frac{i}{2}\sum_{ll'}\sum_{ss'}\hat{c}_{ls}M_{ls,l's'}\hat{c}_{l's'},
\end{equation}
where $M_{ls,l's'} = J_{\left\langle ls,l's' \right\rangle}\hat{u}_{\left\langle ls,l's' \right\rangle}$ and the $1/2$ pre-factor comes from double counting of the bonds. By applying the translational symmetry of the superlattice, $c_{ls}=\frac{1}{\sqrt{N}}\sum_{\mathbf{k}}e^{i\mathbf{k}\cdot\mathbf{r}_{ls}}$, the Hamiltonian becomes
\begin{equation}
    \mathcal{H} = \frac{1}{2}\sum_{\mathbf{k}ss'}\hat{c}_{\mathbf{k}s}^{\dagger}\mathcal{H}_{\mathbf{k}ss'}\hat{c}_{\mathbf{k}s'},
\end{equation}
 where 
\begin{equation}
    \mathcal{H}_{\mathbf{k}ss'} = i\sum_{l}e^{-i\mathbf{k}\cdot\left(\mathbf{r}_{ls}-\mathbf{r}_{l's'}\right)}M_{ls,l's'} = \sum_{nn'}U_{\mathbf{k}sn}E_{\mathbf{k}nn'}U_{\mathbf{k}n's'}^{\dagger}.
\end{equation}
Since $E_{\mathbf{k}}$ is the diagonal matrix of energy eigenvalues for a given $\mathbf{k}$, the Hamiltonian and the energy current operator can be written in terms of the eigenmodes 
\begin{equation}
\mathcal{H} = \sum_{\mathbf{k}n}E_{\mathbf{k}n}\left(\hat{a}_{\mathbf{k}n}^{\dagger}\hat{a}_{\mathbf{k}n}-\frac{1}{2}\right), \, \mathbf{J} = \sum_{\mathbf{k}nn'}\mathbf{J}_{\mathbf{k}nn'}\hat{a}_{\mathbf{k}n}^{\dagger}\hat{a}_{\mathbf{k}n'}.
\end{equation}

One can decompose the $\kappa_{xx}(\omega, T)$ into the  Drude weight ($\omega = 0$) and the dynamical dissipative part as \cite{Metavitsiadis2017,Pidatella2019}
  \begin{align}
  & \kappa_{xx}(\omega, T)=2\pi D (T) \delta(\omega)+ \kappa_{xx}^{\rm reg}(\omega, T),
  \end{align}
  where
  \begin{align}
   D(T) = \frac{1}{ZT^{2}V}\sum_{\substack{ \mathbf{k}nn' \\ E_{\mathbf{k}n}=E_{\mathbf{k}n'}}}e^{-\beta E_{\mathbf{k}n}}|J^{x}_{\mathbf{k}nn'}|^{2},
  \end{align}
  \begin{align}\label{eq:kappa_xx}
   \kappa_{xx}^{\rm reg}(\omega\neq 0, T) = \frac{-2\pi}{ZTV}\sum_{\substack{ \mathbf{k}nn' \\ E_{\mathbf{k}n}\neq E_{\mathbf{k}n'}}}\left(\frac{e^{-\beta E_{\mathbf{k}n'}}-e^{-\beta E_{\mathbf{k}n}}}{E_{\mathbf{k}n'}-E_{\mathbf{k}n}}\right)|J^{x}_{\mathbf{k}nn'}|^{2}\delta[\omega-(E_{\mathbf{k}n}-E_{\mathbf{k}n'})]\,.
  \end{align}
 The Drude weight  $D\equiv D(T)$ is a measure of the non-dissipative contribution to the heat flow, and a non-zero $D$ corresponds to a perfect conductor. 
 The dynamical part  $\kappa_{xx}^{\rm reg}(\omega, T)$ corresponds to  the dissipative contribution to  the heat flow. If both  $D=0$ and $\kappa_{xx}^{\rm reg}(\omega\rightarrow 0, T)=0$, the system is an insulator. For the shortness of notations,  in the following we  remove the superscript ${\rm reg}$ and  denote  the dynamical part of the thermal conductivity as $\kappa_{xx}(\omega, T)$.

We first examine the Drude weight in static flux backgrounds. In the pure model with ground-state zero-flux sector, the energy current operator commutes with the Hamiltonian, and thus there are no off-diagonal matrix elements that contribute to the finite-frequency thermal conductivity \cite{Nasu2017, Metavitsiadis2017, Pidatella2019}. In this case, the Drude weight is finite for $T > 0$ and the peak locates around $T \sim J$, as shown in Fig.~\ref{fig:Drude_weight} (a). However, in the presence of weak disorder, e.g. when having 2$\%$ of true vacancies in the bound-flux sector, the finite-temperature Drude weight diminishes. The finite-size trend in Fig.~\ref{fig:Drude_weight} (b) of the peak value suggests that the Drude weight completely vanishes in the thermodynamic limit, indicating that the thermal transport is no longer ballistic even in the presence of the weak disorder. For the case of random fluxes (not shown  in Fig.~\ref{fig:Drude_weight} (a)), the disorder is already strong enough to destroy the Drude weight for finite-size systems.

\begin{figure*}
     \includegraphics[width=1.0\textwidth]{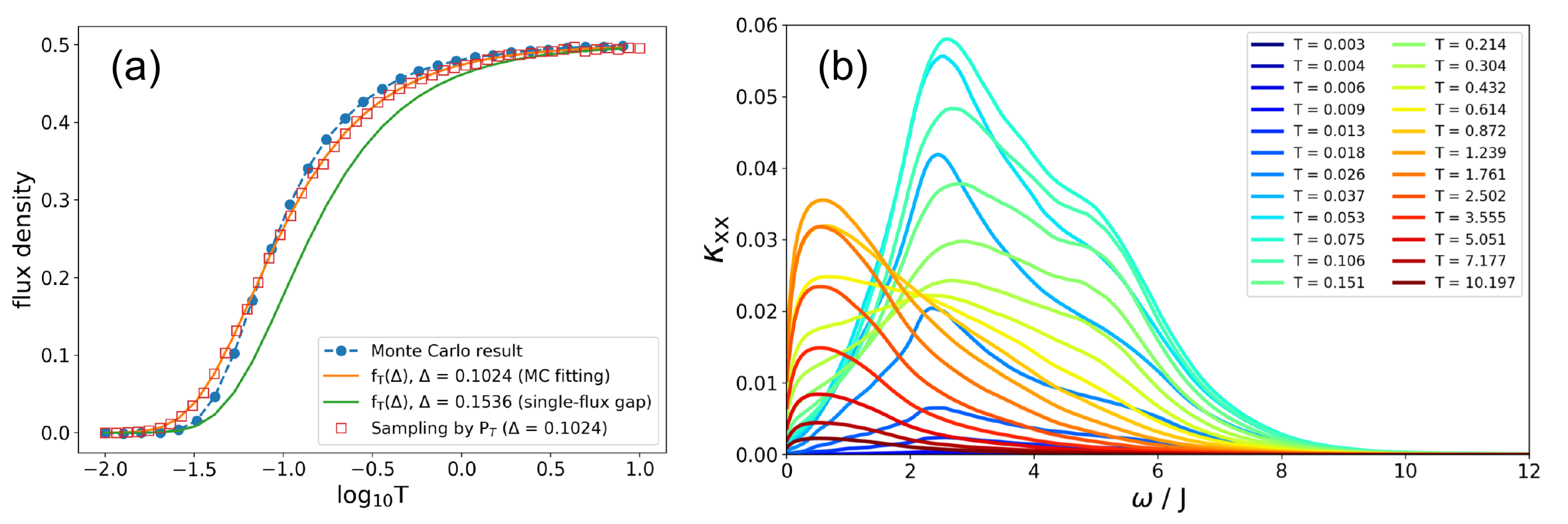}
     \caption{\label{fig:thermal_flux} Longitudinal thermal conductivity in the thermal-flux sector. (a) Based on the $L = 32$ Monte Carlo result \cite{Feng2021}, the temperature dependence of the flux density can be fitted by a Fermi-Dirac function. The fitted flux gap $\Delta  = 0.1024$ is then used in the flipping probability (Eq.~\ref{Eq:probability}) of the $\hat{u}$ variables. (b) Frequency- and temperature-dependence of $\kappa_{xx}$ in the pristine Kitaev model. The calculations are done for a superlattice $N = 100$ and cluster size $2L^{2} = 800$, with average over 50 disorder samples.}
\end{figure*}

\begin{figure*}
     \includegraphics[width=1.0\textwidth]{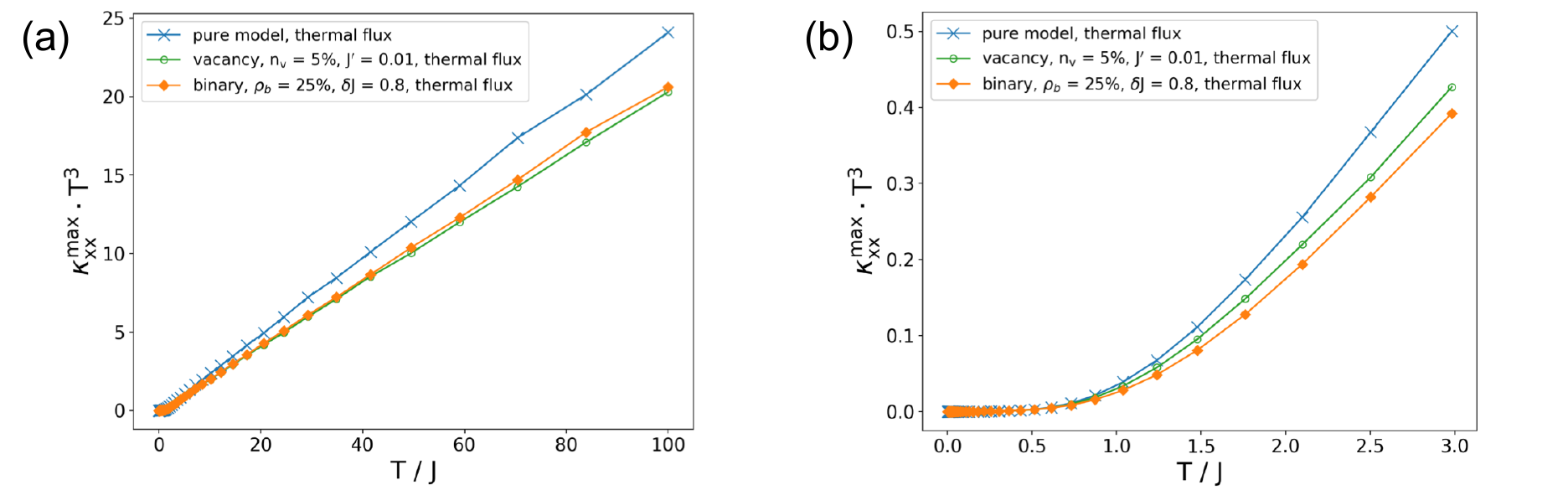}
     \caption{\label{fig:kappa_xx_peak} The temperature dependence of $\kappa^{\mathrm{max}}_{xx} (\omega)$ peak values. (a) At high temperatures, as the DOS is flat and dominated by the random-flux background,  the temperature dependence of  $\kappa_{xx}^{\mathrm{max}}$ follows the behavior of Eq.~(\ref{Eq:kappa_xx_peak}). (b)  At temperatures below $T \sim J$, the temperature dependence deviates from Eq.~(\ref{Eq:kappa_xx_peak}) since the thermal-flux density decreases and the assumption of a flat DOS is no longer valid.}
\end{figure*}

In order to incorporate the effect of thermal flux excitations, we assign random-flux sector with specific flux density for each temperature. The temperature-dependent flux density function is extracted from the Monte Carlo result of the pure Kitaev model \cite{Feng2021}, as shown in Fig.~\ref{fig:thermal_flux} (a). We first fit the data by a Fermi-Dirac function
\begin{equation}
    f_{T}(\Delta) = \frac{1}{e^{\Delta/T}+1},
\end{equation}
where we obtain $\Delta \sim 0.1024$. Note that this flux gap is smaller than the single-flux gap $0.1536$ reported in Ref.~\onlinecite{Kitaev2006} due to the interactions between fluxes in a finite system \cite{Feng2020}. Instead of sampling the thermal fluxes by Monte Carlo simulations, we take the disorder average over the typical flux sectors that governed by the above distribution. As demonstrated in Ref.~\onlinecite{Gabor2019}, this can be done by assigning a temperature-dependent flipping probability $P_{T}$ to each link variable $\hat{u}$ in the system
\begin{equation}\label{Eq:probability}
    P_{T} = \frac{1-\left[1-2f_{T}(\Delta)\right]^{1/6}}{2}.
\end{equation}
We term the flux-sectors given by this random flipping probability the \textit{thermal-flux sectors}. In Fig.~\ref{fig:thermal_flux} (a) we verify that the above probability leads to precise sampling in the thermal flux density.
Note that in disordered systems, this approximation can deviate from the true thermal-flux distribution, especially at very low temperatures. The flipping probability is based on the fitted flux gap of pure systems, such that the quench disorder effect to the flux gap is not included. Therefore, at temperatures lower than $\Delta$, the systems are mostly in the zero-flux sector.

In Fig.~\ref{fig:thermal_flux} (b), we show the frequency-dependent longitudinal thermal conductivity $\kappa_{xx}(\omega, T)$ for a pristine Kitaev model. The results are disorder averaged, but in this case the randomness only comes from the thermal-flux sector at high temperatures. At very low temperatures, on the other hand, the system is mostly  in the zero-flux sector, such that the energy current operator is still a conserved quantity and its time-correlation function is actually time independent \cite{Zotos1997}. Therefore, the off-diagonal matrix elements $|J^{x}_{\mathbf{k}nn^{\prime}}|$ in Eq.~(\ref{eq:kappa_xx}) dominate the low-temperature behavior, and the non-zero frequency contribution to $\kappa_{xx} (\omega, T)$ is negligible. As temperature increases ($0 < T 	\lesssim 0.1$), a broad peak centering around $\omega / J \sim 2.5$ is gradually formed. The peak position corresponds to the van Hove singularity in the density of states. For temperatures higher than the energy scale of thermal flux gap ($T \gtrsim 0.1$), the van Hove singularity vanishes and the DOS is flattened.  Therefore, in the high-temperature limit, the temperature dependence of the peak value  $\kappa^{\mathrm{max}}_{xx}(\omega, T)$  can be estimated by
\begin{equation}\label{Eq:kappa_xx_peak}
    \kappa_{xx}^{\mathrm{max}}(\omega, T) \sim \sum_{\mathbf{k}nn'} \frac{2}{\omega T}e^{-\frac{\Omega_{\mathbf{k}nn'}}{2T}}\sinh{\left( \frac{\omega}{2T}\right )} \sim \sum_{\mathbf{k}nn'}\frac{1}{T^{2}}\left( 1-\frac{\Omega_{\mathbf{k}nn'}}{2T}\right),
\end{equation}
where  $\omega = E_{\mathbf{k}n^{\prime}} - E_{\mathbf{k}n}$ and we define $\Omega_{\mathbf{k}nn'} \equiv E_{\mathbf{k}n^{\prime}} + E_{\mathbf{k}n}$. In Fig.~\ref{fig:kappa_xx_peak}, we demonstrate this temperature dependence of $\kappa_{xx}^{\mathrm{max}}$ by plotting  $\kappa_{xx}^{\mathrm{max}}T^3$ versus $T/J$. In Fig.~\ref{fig:kappa_xx_peak} (a) we show this dependence (blue line for the pristine model) for all temperatures and  in (b) we focus only  its low-temperature part. At high temperatures, the DOS is flat and dominated by the random-flux background, such that $\kappa_{xx}^{\mathrm{max}}$ follows the behavior of Eq.~(\ref{Eq:kappa_xx_peak}), and as a result $\kappa_{xx}^{\mathrm{max}}T^3$ is linear in $T/J$.
When lowering the temperatures (see Fig.~\ref{fig:kappa_xx_peak}(b)), the curve deviates from the linear relationship around $T \sim J$,  below which the thermal fluxes are less proliferated and  the system is not yet in the random-flux sector.

\begin{figure*}
     \includegraphics[width=1.0\textwidth]{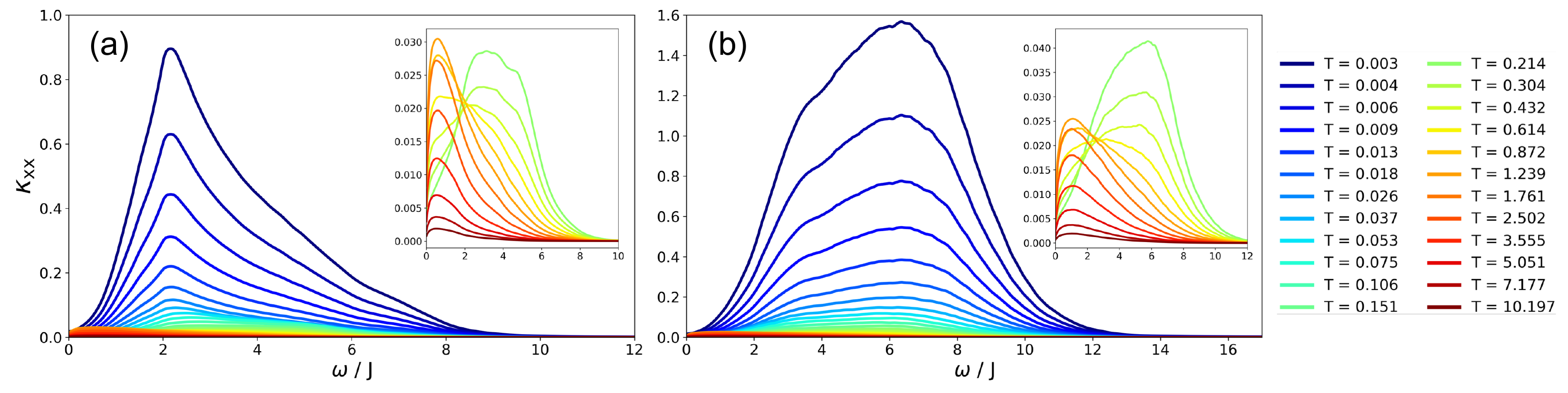}
     \caption{\label{fig:spectrum_disorder} Frequency- and temperature-dependence of the longitudinal thermal conductivity in the presence of (a) 5\% quasivacancy with $J^{\prime} = 0.01$, and (b) binary bond disorder with $\rho_{b} = 25\%$ and $\delta J = 0.8$. The insets show the results for $T > 0.2$. The calculations are done for a superlattice $N = 100$ and cluster size $2L^{2} = 800$, with average over 50 disorder samples.}
\end{figure*}

In the presence of disorder (see Fig.~\ref{fig:spectrum_disorder}), the finite-frequency contribution to the longitudinal thermal conductivity is significant even at low temperatures. As we discussed in the pristine case, the matrix elements of the energy current operator have predominant effect in this region, and the quenched disorders provide notable contributions to the off-diagonal elements even without the thermal-flux disorder. This is consistent with the previous observation in the Drude weight that even weak disorder (small concentration of vacancies or quasivacancies) can turn the system into a dissipative thermal conductor. The  thermal conductivity spectrum for the systems with  vacancies shown in Fig.~\ref{fig:spectrum_disorder} (a)  displays a strongly temperature-dependent behavior which
can be understood from the behavior of the  corresponding DOS presented in Fig.~\ref{fig:DOS_IPR} (c) and (k).  At low temperatures, the flux sector can be considered as a static bound-flux sector  and thus the characteristic behavior of DOS corresponds to Fig.~\ref{fig:DOS_IPR} (c). There we see that the DOS as a function of energy  shows extremely non-monotonous behavior: at  very low energies it shows a  power-law divergence, then rapidly decreases with increasing energy and then  grows again towards the van Hove singularity at  $E\sim J$. At high  temperatures, the flux sector  is well described by the random-flux sector  and, as we can see from Fig.~\ref{fig:DOS_IPR} (k), the DOS is flatten over the whole bandwidth.
Following the trend of the DOS behaviour, thermal conductivity spectrum changes significantly with temperature. 
 At low temperatures, it has a characteristic triangular shape  with a very large value  at the peak, coming predominantly from  high DOS of the Majorana fermions on the vacancy-induced low-energy states in the bound-flux sector.  With increasing temperature, the peak value decreases   rapidly as a result  of decreasing  DOS  at energy above the pileup region.  The inset  in Fig.~\ref{fig:spectrum_disorder} (a) shows that at temperatures above  the pileup region the thermal conductivity spectrum is very similar to the one in the pristine model (see Fig.~\ref{fig:thermal_flux} (b)).

The spectrum in Fig.~\ref{fig:spectrum_disorder} (b) corresponds to the strong binary bond-disorder. 
 Its shape and temperature dependence can again be understood  from the behavior of the corresponding DOS shown in Fig.~\ref{fig:DOS_IPR} (f) and (l). At low temperatures, the system is  in the zero-flux sector and its characteristic DOS   is shown Fig.~\ref{fig:DOS_IPR} (f), where we see that it does not have neither  a 
 low-energy  power-law divergence nor the van Hove singularity. Instead, the strong disorder significantly flattens the DOS  both in the  zero-flux (Fig.~\ref{fig:DOS_IPR} (f)) and in the random-flux sectors (Fig.~\ref{fig:DOS_IPR} (l)). As result of this flatten DOS, the spectrum is 
  smooth and rounded already at low temperatures. Again,  at higher temperatures the behavior of the spectrum (see inset in Fig.~\ref{fig:spectrum_disorder} (b)) is  similar to the pristine case (Fig.~\ref{fig:thermal_flux} (b)).  For both the site- and bond-disorder case, the temperature behavior  of the peak value $\kappa_{xx}^{\mathrm{max}}$ is qualitatively similar to the pristine case (see Fig.~\ref{fig:kappa_xx_peak}).

\begin{figure*}
     \includegraphics[width=1.0\textwidth]{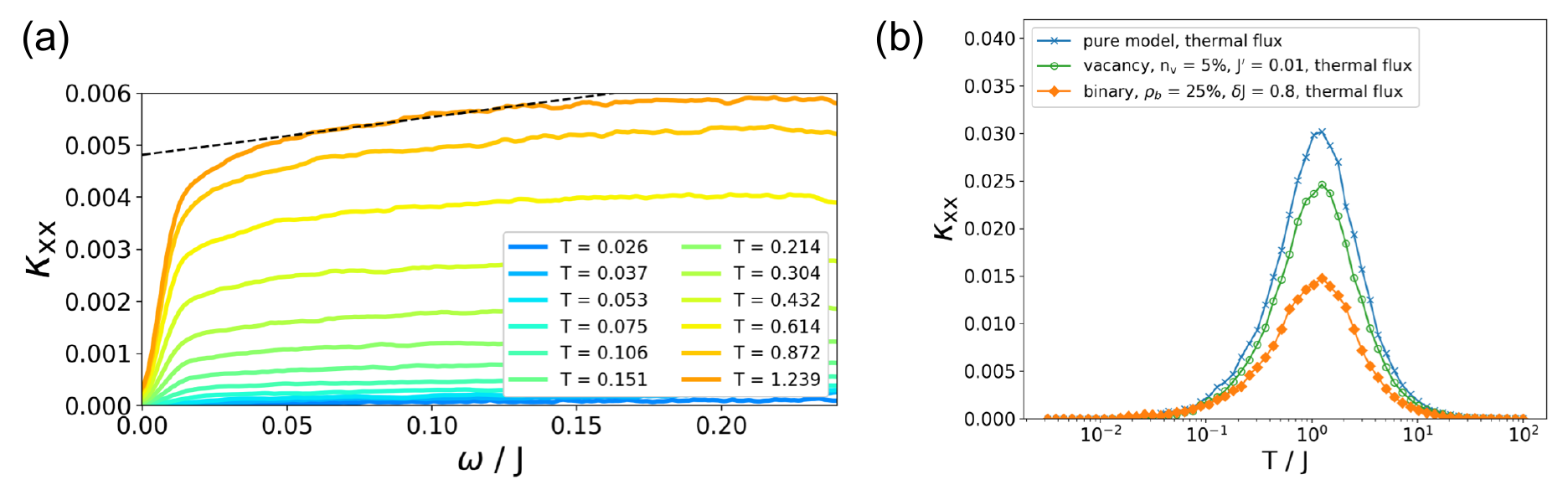}
     \caption{\label{fig:conductivity_thermal_flux} Temperature-dependence of $\kappa_{xx}$. (a) The $\omega \to 0$ contribution at each temperature is extrapolated by linear regression of $\kappa_{xx}(\omega)$ in the range of $0.05 \lesssim \omega \lesssim 0.15$. (b) The result of $\kappa_{xx}$ shows a single-peak structure around $T\sim J$, which is suppressed by both site- or bond-disorder.} 
\end{figure*}

Finally, in order to see the overall temperature dependence of $\kappa_{xx}$, we extrapolate the low-frequency part and obtain $\kappa_{xx}(T) = \lim_{\omega \to 0}\kappa_{xx}(T,\omega)$. In Fig.~\ref{fig:conductivity_thermal_flux} (b), the maximum of $\kappa_{xx}$ happens around the Kitaev-coupling energy scale, and the presence of disorder indeed suppress the longitudinal thermal conductivity. This result is consistent with the recent Monte Carlo study on disordered Kitaev spin liquid by Nasu \etal ~\cite{Motome2020}. Unlike the specific heat result that shows two crossovers which represent flux excitations and itinerant Majorana fermion excitations respectively, in the longitudinal thermal conductivity only the latter contributes to heat transport notably. However, those thermally proliferated fluxes saturate to the density $P_{T} = 1/2$ as $T/J \gtrsim 1$ (Fig.~\ref{fig:thermal_flux} (a)), and thus scatter the itinerant Majoran fermions and reduce the longitudinal thermal conductivity. With quenched disorders, the peak around $T \approx J$ is further reduced by the presence of random scatterers.

\section{Conclusion}
\label{Conclusion}

In this paper, we studied the disorder and localization effects in the two-dimensional Kitaev QSL. In this system, the ideas of quantum spin liquid and one-body localization, which both can be traced back to the seminal works of P. W. Anderson, intertwined through the concept of spin fractionalization. The excitation spectrum is described by free Majorana fermions with graphene-like density of states, along with the gapped local $Z_{2}$ gauge fluxes. In a static flux background, the non-interacting fermionic Hamiltonian provides the exact solvability of the model and thus the accessibility to numerically study of the localization effects in the presence of disorder. Interestingly, even without vacancies or random bonds, the thermally proliferated fluxes in the pristine model engender the intrinsic disorder and suppress the ballistic thermal transport \cite{Metavitsiadis2017}. By examining various kinds of extrinsic disorder, we show that the density of states and localization behaviors can be diverse. For instance, vacancies induce a power-law divergence of DOS in the low-energy limit. The states in this pileup are more localized than most of the other states of the system, which can be shown by the distribution of inverse participation ratio.
The localization of low-energy states is particularly strong  when time-reversal symmetry is broken by the three-spin interaction term $h$ that imitates an effect of the external magnetic field. 
By contrast, in the Kitaev spin liquid with purely bond disorder, the localization does not happen for the low-energy states  but for the Majorana fermion eigenstates at the high-energy edge, known as the Lifshitz tails. By considering the temperature-dependent density of fluxes (thermal-flux sector), we also showed that  both types of quenched disorder tend to reduce the longitudinal thermal conductivity at finite temperatures. 

Motivated by the peculiar power-law divergence of specific heat $C/T$ in the Kitaev QSL candidate H$_{3}$LiIr$_{2}$O$_{6}$ \cite{Kitagawa2018}, the disorder-induced pileup in the low-energy DOS in Kitaev QSL becomes an interesting topic to explore. In Ref.~\onlinecite{Knolle2019}, it was shown that a strong bond disorder with random-flux sector gives rise to a similar power-law behavior. On the other hand, in Ref.~\onlinecite{Kao2020} it was shown that this pileup can alternatively be ascribable to the small amount of vacancies or quasivacancies in the system.  Basically, both types of disorder lead to the accumulation of low-energy states which can be observed in the specific heat measurements.
In this work, however, by comparing the distributions of IPR we discovered that the localization behaviors of the low-energy states are distinguishable in the two scenarios. While the vacancy-induced low-energy states are more localized than the high-energy states, the low-energy states in the case of binary disorder and random flux are more delocalized than the high-energy states. The distinction of localization behaviors is more obvious when the three-spin interaction $h$ is applied.
A recent work of Monte Carlo simulations suggests that the low-temperature plateau of thermal Hall conductivity is more robust in the bond-disordered Kitaev QSL (with box distribution) than the site-disordered counterpart \cite{Motome2020}. Therefore, whether the two scenarios of DOS pileups mentioned above can lead to distinct responses in the thermal Hall conductivity is worth further investigations, and may shed light on the nature of low-energy physics in H$_{3}$LiIr$_{2}$O$_{6}$. 
In addition, recent experimental findings in the measurements of nuclear magnetic resonance and longitudinal thermal conductivity in pristine $\alpha$-RuCl$_{3}$ and diluted compound $\alpha$-Ru$_{1-x}$Ir$_{x}$Cl$_{3}$ reveal that defect-induced low-energy excitations may play an important role \cite{Hentrich2020, Baek2020, Do2020}. Thus, the future investigation on disordered Kitaev spin liquid may also lead us to a deeper understanding of field-induced quantum spin liquids. 
 
  \vspace*{0.3cm} 
\noindent{\it  Acknowledgments:} 
We thank  Kexin Feng, G\'abor B. Hal\'asz, Johannes Knolle, Roderich Moessner and Masahiko Yamada for valuable discussions.
Our work was supported by the  National Science Foundation  under Award No.\ DMR-1929311.   N.B.P. acknowledges the hospitality of Kavli Institute for Theoretical Physics and
 the National Science Foundation under Grant No. NSF PHY-1748958.
 
\appendix*
\section{Antiferromagnetic random couplings}

\begin{figure*}
     \includegraphics[width=1.0\textwidth]{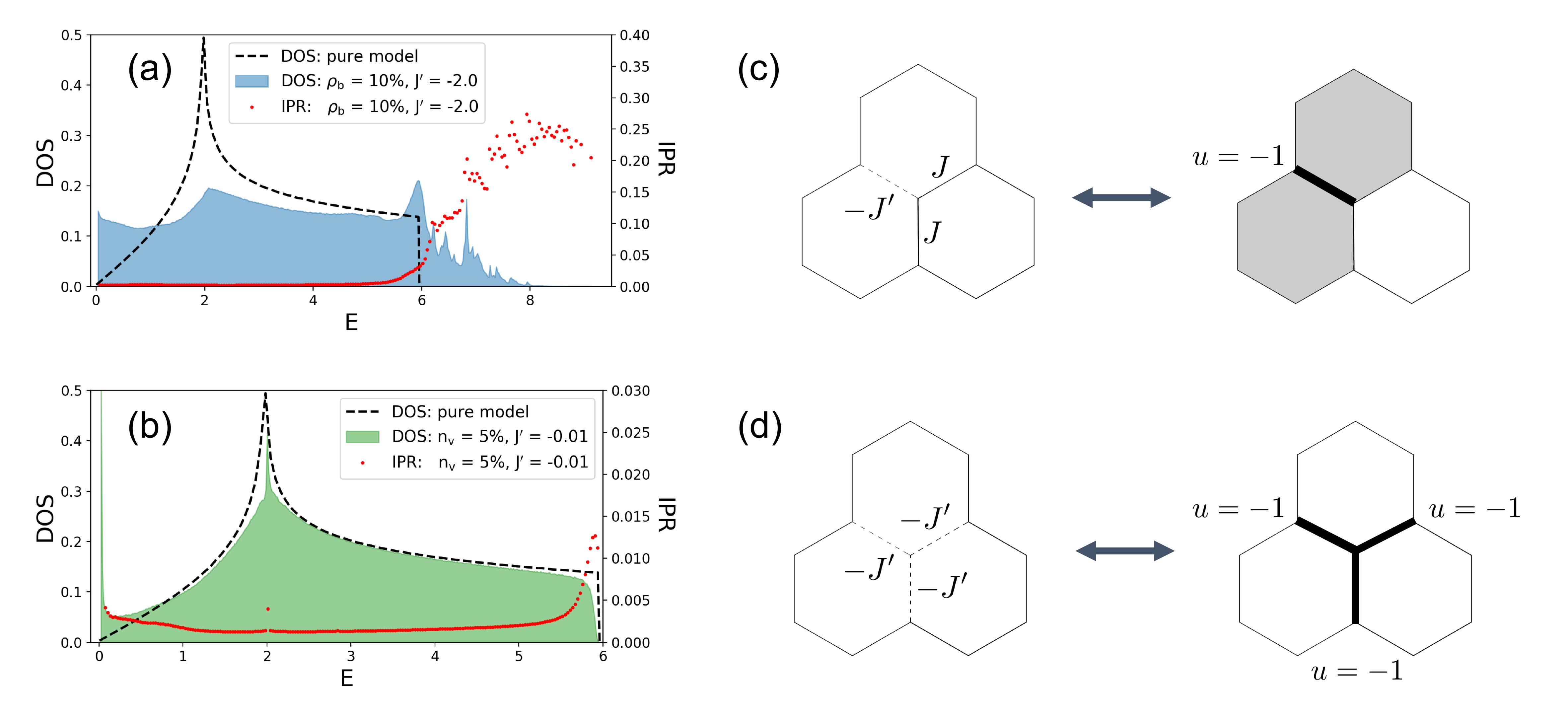}
     \caption{\label{fig:appendix} Antiferromagnetic random couplings $J^{\prime} < 0$. (a)-(b): DOS and IPR for the system with random bonds and quasivacancies, respectively. The details of the calculation are the same as in Fig.~\ref{fig:DOS_IPR}. (c)-(d): The equivalence between the change of sign in $J^{\prime}$ and local flip of $u$ variables, for the geometry of a random bond and a quasivacancy. The flipped u variables are depicted as black thick lines and the fluxes ($W = -1$) are shown as gray plaquettes.} 
\end{figure*}

Throughout this work, we consider the ferromagnetic interaction for both normal couplings ($J > 0$) and random couplings ($J^{\prime} > 0$). In the Kitaev honeycomb model, the overall sign will not change the physics because it corresponds to a simple gauge transformation of the Majorana hopping Hamiltonian. However, if we keep the normal couplings ferromagnetic but turn the random couplings into antiferromagnetic, the spectrum and density of states may change drastically. In this appendix, we briefly summarize the effect of antiferromagnetic random couplings on the exactly-solvable Kitaev honeycomb model.

Here we consider two scenarios of turning random $J^{\prime}$ into antiferromagnetic couplings: random bonds with $\rho_{b} = 10\%$ and $J^{\prime} = -2.0$, corresponding to Fig.~\ref{fig:DOS_IPR} (e), and the case of quasivacancies with $n_{v} = 5\%$ and $J^{\prime} = -0.01$, corresponding to Fig.~\ref{fig:DOS_IPR} (c).  The calculation details are the same as in Fig.~\ref{fig:DOS_IPR}, with the only difference that $J^{\prime}$ becomes negative value on all the random couplings. In Fig.~\ref{fig:appendix} (a)-(b), The numerical results show that in the former case of doped random bonds, the DOS is changed drastically by the transformation into negative $J^{\prime}$, while in the latter case of quasivacncies, the DOS remains the same. 

The difference between these two cases can be understood easily by the following argument. In the nearest-neighbor Majorana hopping Hamiltonian, the coupling strength $J^{\prime}$ and the link variable $u$ always appear as a product,
namely, each hopping term in the Hamiltonian is in the form of $J^{\prime}_{\langle ij \rangle}\hat{u}_{\langle ij \rangle}\hat{c}_{i}\hat{c}_{j}$, such that the transformation from ferro- to antiferromagnetic interaction is equivalent to the local flip of $u$ variable:
\begin{equation}
    J^{\prime} \to -J^{\prime} \quad \Longleftrightarrow \quad u = +1 \to u = -1.
\end{equation}

Therefore, these local flips may or may not change the flux sector of the system, depending on the real-space distribution of random bonds $J^{\prime}$. In Fig.~\ref{fig:appendix} (c)-(d), we demonstrate that the sign change of  $J^{\prime}$ has an equivalent effect to  the local flip of u variables.
 
In the case of random bonds, transformation from $J^{\prime}$ to $-J^{\prime}$ is equivalent as keeping $J^{\prime}$ ferromagnetic but flipping $u$ on the same bond, leading to the creation of a flux pair. As a result, at finite density of random couplings, the flux sector will become similar to the random-flux sector as in Fig.~\ref{fig:DOS_IPR} (j)-(l), and the flux configuration is determined by the real-space distribution of the antiferromagnetic bonds.
 
On the other hand, in the case of quasivacancy, random couplings $J^{\prime}$ always appear as trimers, such that the flux sector remains intact before and after the transformation. This explains why the sign change of $J^{\prime}$ has no effect on the density of states. In summary, we show that the random antiferromagnetic couplings can lead to different behaviors for the bond-disordered and site-diluted Kitaev spin liquid based on the local geometry of the random bonds, providing an interesting future perspective for the disorder effect in Kitaev materials.

\bibliographystyle{apsrev4-1}
\bibliography{DisorderKitaev.bib}

\begin{thebibliography}{93}%
\makeatletter
\providecommand \@ifxundefined [1]{%
 \@ifx{#1\undefined}
}%
\providecommand \@ifnum [1]{%
 \ifnum #1\expandafter \@firstoftwo
 \else \expandafter \@secondoftwo
 \fi
}%
\providecommand \@ifx [1]{%
 \ifx #1\expandafter \@firstoftwo
 \else \expandafter \@secondoftwo
 \fi
}%
\providecommand \natexlab [1]{#1}%
\providecommand \enquote  [1]{``#1''}%
\providecommand \bibnamefont  [1]{#1}%
\providecommand \bibfnamefont [1]{#1}%
\providecommand \citenamefont [1]{#1}%
\providecommand \href@noop [0]{\@secondoftwo}%
\providecommand \href [0]{\begingroup \@sanitize@url \@href}%
\providecommand \@href[1]{\@@startlink{#1}\@@href}%
\providecommand \@@href[1]{\endgroup#1\@@endlink}%
\providecommand \@sanitize@url [0]{\catcode `\\12\catcode `\$12\catcode
  `\&12\catcode `\#12\catcode `\^12\catcode `\_12\catcode `\%12\relax}%
\providecommand \@@startlink[1]{}%
\providecommand \@@endlink[0]{}%
\providecommand \url  [0]{\begingroup\@sanitize@url \@url }%
\providecommand \@url [1]{\endgroup\@href {#1}{\urlprefix }}%
\providecommand \urlprefix  [0]{URL }%
\providecommand \Eprint [0]{\href }%
\providecommand \doibase [0]{http://dx.doi.org/}%
\providecommand \selectlanguage [0]{\@gobble}%
\providecommand \bibinfo  [0]{\@secondoftwo}%
\providecommand \bibfield  [0]{\@secondoftwo}%
\providecommand \translation [1]{[#1]}%
\providecommand \BibitemOpen [0]{}%
\providecommand \bibitemStop [0]{}%
\providecommand \bibitemNoStop [0]{.\EOS\space}%
\providecommand \EOS [0]{\spacefactor3000\relax}%
\providecommand \BibitemShut  [1]{\csname bibitem#1\endcsname}%
\let\auto@bib@innerbib\@empty
\bibitem [{\citenamefont {Anderson}(1973)}]{anderson1973resonating}%
  \BibitemOpen
  \bibfield  {author} {\bibinfo {author} {\bibfnamefont {P.~W.}\ \bibnamefont
  {Anderson}},\ }\href
  {https://www.sciencedirect.com/science/article/pii/0025540873901670}
  {\bibfield  {journal} {\bibinfo  {journal} {Mater. Res. Bull.}\ }\textbf
  {\bibinfo {volume} {8}},\ \bibinfo {pages} {153} (\bibinfo {year}
  {1973})}\BibitemShut {NoStop}%
\bibitem [{\citenamefont {Huse}\ and\ \citenamefont {Elser}(1988)}]{Huse88}%
  \BibitemOpen
  \bibfield  {author} {\bibinfo {author} {\bibfnamefont {D.~A.}\ \bibnamefont
  {Huse}}\ and\ \bibinfo {author} {\bibfnamefont {V.}~\bibnamefont {Elser}},\
  }\href {\doibase 10.1103/PhysRevLett.60.2531} {\bibfield  {journal} {\bibinfo
   {journal} {Phys. Rev. Lett.}\ }\textbf {\bibinfo {volume} {60}},\ \bibinfo
  {pages} {2531} (\bibinfo {year} {1988})}\BibitemShut {NoStop}%
\bibitem [{\citenamefont {Laughlin}(1983)}]{Laughlin1983}%
  \BibitemOpen
  \bibfield  {author} {\bibinfo {author} {\bibfnamefont {R.~B.}\ \bibnamefont
  {Laughlin}},\ }\href {\doibase 10.1103/PhysRevLett.50.1395} {\bibfield
  {journal} {\bibinfo  {journal} {Phys. Rev. Lett.}\ }\textbf {\bibinfo
  {volume} {50}},\ \bibinfo {pages} {1395} (\bibinfo {year}
  {1983})}\BibitemShut {NoStop}%
\bibitem [{\citenamefont {Kalmeyer}\ and\ \citenamefont
  {Laughlin}(1987)}]{Kalmeyer1987}%
  \BibitemOpen
  \bibfield  {author} {\bibinfo {author} {\bibfnamefont {V.}~\bibnamefont
  {Kalmeyer}}\ and\ \bibinfo {author} {\bibfnamefont {R.~B.}\ \bibnamefont
  {Laughlin}},\ }\href {\doibase 10.1103/PhysRevLett.59.2095} {\bibfield
  {journal} {\bibinfo  {journal} {Phys. Rev. Lett.}\ }\textbf {\bibinfo
  {volume} {59}},\ \bibinfo {pages} {2095} (\bibinfo {year}
  {1987})}\BibitemShut {NoStop}%
\bibitem [{\citenamefont {Lee}(2008)}]{Lee2008}%
  \BibitemOpen
  \bibfield  {author} {\bibinfo {author} {\bibfnamefont {P.~A.}\ \bibnamefont
  {Lee}},\ }\href {https://science.sciencemag.org/content/321/5894/1306}
  {\bibfield  {journal} {\bibinfo  {journal} {Science}\ }\textbf {\bibinfo
  {volume} {321}},\ \bibinfo {pages} {1306} (\bibinfo {year}
  {2008})}\BibitemShut {NoStop}%
\bibitem [{\citenamefont {Balents}(2010)}]{Balents2010}%
  \BibitemOpen
  \bibfield  {author} {\bibinfo {author} {\bibfnamefont {L.}~\bibnamefont
  {Balents}},\ }\href {https://www.nature.com/articles/nature08917} {\bibfield
  {journal} {\bibinfo  {journal} {Nature}\ }\textbf {\bibinfo {volume} {464}},\
  \bibinfo {pages} {199} (\bibinfo {year} {2010})}\BibitemShut {NoStop}%
\bibitem [{\citenamefont {Savary}\ and\ \citenamefont
  {Balents}(2017{\natexlab{a}})}]{Savary2016}%
  \BibitemOpen
  \bibfield  {author} {\bibinfo {author} {\bibfnamefont {L.}~\bibnamefont
  {Savary}}\ and\ \bibinfo {author} {\bibfnamefont {L.}~\bibnamefont
  {Balents}},\ }\href {http://stacks.iop.org/0034-4885/80/i=1/a=016502}
  {\bibfield  {journal} {\bibinfo  {journal} {Rep. Prog. Phys.}\ }\textbf
  {\bibinfo {volume} {80}},\ \bibinfo {pages} {016502} (\bibinfo {year}
  {2017}{\natexlab{a}})}\BibitemShut {NoStop}%
\bibitem [{\citenamefont {Zhou}\ \emph {et~al.}(2017)\citenamefont {Zhou},
  \citenamefont {Kanoda},\ and\ \citenamefont {Ng}}]{Zhou2017}%
  \BibitemOpen
  \bibfield  {author} {\bibinfo {author} {\bibfnamefont {Y.}~\bibnamefont
  {Zhou}}, \bibinfo {author} {\bibfnamefont {K.}~\bibnamefont {Kanoda}}, \ and\
  \bibinfo {author} {\bibfnamefont {T.-K.}\ \bibnamefont {Ng}},\ }\href
  {\doibase 10.1103/RevModPhys.89.025003} {\bibfield  {journal} {\bibinfo
  {journal} {Rev. Mod. Phys.}\ }\textbf {\bibinfo {volume} {89}},\ \bibinfo
  {pages} {025003} (\bibinfo {year} {2017})}\BibitemShut {NoStop}%
\bibitem [{\citenamefont {Knolle}\ and\ \citenamefont
  {Moessner}(2019)}]{KnolleMoessner2019}%
  \BibitemOpen
  \bibfield  {author} {\bibinfo {author} {\bibfnamefont {J.}~\bibnamefont
  {Knolle}}\ and\ \bibinfo {author} {\bibfnamefont {R.}~\bibnamefont
  {Moessner}},\ }\href {\doibase 10.1146/annurev-conmatphys-031218-013401}
  {\bibfield  {journal} {\bibinfo  {journal} {Annu. Rev. Condens. Matter
  Phys.}\ }\textbf {\bibinfo {volume} {10}},\ \bibinfo {pages} {451} (\bibinfo
  {year} {2019})}\BibitemShut {NoStop}%
\bibitem [{\citenamefont {Broholm}\ \emph {et~al.}(2020)\citenamefont
  {Broholm}, \citenamefont {Cava}, \citenamefont {Kivelson}, \citenamefont
  {Nocera}, \citenamefont {Norman},\ and\ \citenamefont
  {Senthil}}]{Broholm2020}%
  \BibitemOpen
  \bibfield  {author} {\bibinfo {author} {\bibfnamefont {C.}~\bibnamefont
  {Broholm}}, \bibinfo {author} {\bibfnamefont {R.~J.}\ \bibnamefont {Cava}},
  \bibinfo {author} {\bibfnamefont {S.~A.}\ \bibnamefont {Kivelson}}, \bibinfo
  {author} {\bibfnamefont {D.~G.}\ \bibnamefont {Nocera}}, \bibinfo {author}
  {\bibfnamefont {M.~R.}\ \bibnamefont {Norman}}, \ and\ \bibinfo {author}
  {\bibfnamefont {T.}~\bibnamefont {Senthil}},\ }\href
  {https://science.sciencemag.org/content/367/6475/eaay0668} {\bibfield
  {journal} {\bibinfo  {journal} {Science}\ }\textbf {\bibinfo {volume} {367}}
  (\bibinfo {year} {2020})}\BibitemShut {NoStop}%
\bibitem [{\citenamefont {Takagi}\ \emph {et~al.}(2019)\citenamefont {Takagi},
  \citenamefont {Takayama}, \citenamefont {Jackeli}, \citenamefont
  {Khaliullin},\ and\ \citenamefont {Nagler}}]{Takagi2019}%
  \BibitemOpen
  \bibfield  {author} {\bibinfo {author} {\bibfnamefont {H.}~\bibnamefont
  {Takagi}}, \bibinfo {author} {\bibfnamefont {T.}~\bibnamefont {Takayama}},
  \bibinfo {author} {\bibfnamefont {G.}~\bibnamefont {Jackeli}}, \bibinfo
  {author} {\bibfnamefont {G.}~\bibnamefont {Khaliullin}}, \ and\ \bibinfo
  {author} {\bibfnamefont {S.~E.}\ \bibnamefont {Nagler}},\ }\href {\doibase
  10.1038/s42254-019-0038-2} {\bibfield  {journal} {\bibinfo  {journal} {Nat.
  Rev. Phys.}\ }\textbf {\bibinfo {volume} {1}},\ \bibinfo {pages} {264}
  (\bibinfo {year} {2019})}\BibitemShut {NoStop}%
\bibitem [{\citenamefont {Kitaev}(2006)}]{Kitaev2006}%
  \BibitemOpen
  \bibfield  {author} {\bibinfo {author} {\bibfnamefont {A.}~\bibnamefont
  {Kitaev}},\ }\href {\doibase http://dx.doi.org/10.1016/j.aop.2005.10.005}
  {\bibfield  {journal} {\bibinfo  {journal} {Annals of Physics}\ }\textbf
  {\bibinfo {volume} {321}},\ \bibinfo {pages} {2 } (\bibinfo {year}
  {2006})}\BibitemShut {NoStop}%
\bibitem [{\citenamefont {Norman}(2016)}]{Norman2016}%
  \BibitemOpen
  \bibfield  {author} {\bibinfo {author} {\bibfnamefont {M.~R.}\ \bibnamefont
  {Norman}},\ }\href {\doibase 10.1103/RevModPhys.88.041002} {\bibfield
  {journal} {\bibinfo  {journal} {Rev. Mod. Phys.}\ }\textbf {\bibinfo {volume}
  {88}},\ \bibinfo {pages} {041002} (\bibinfo {year} {2016})}\BibitemShut
  {NoStop}%
\bibitem [{\citenamefont {Shimizu}\ \emph {et~al.}(2003)\citenamefont
  {Shimizu}, \citenamefont {Miyagawa}, \citenamefont {Kanoda}, \citenamefont
  {Maesato},\ and\ \citenamefont {Saito}}]{Shimizu2003}%
  \BibitemOpen
  \bibfield  {author} {\bibinfo {author} {\bibfnamefont {Y.}~\bibnamefont
  {Shimizu}}, \bibinfo {author} {\bibfnamefont {K.}~\bibnamefont {Miyagawa}},
  \bibinfo {author} {\bibfnamefont {K.}~\bibnamefont {Kanoda}}, \bibinfo
  {author} {\bibfnamefont {M.}~\bibnamefont {Maesato}}, \ and\ \bibinfo
  {author} {\bibfnamefont {G.}~\bibnamefont {Saito}},\ }\href {\doibase
  10.1103/PhysRevLett.91.107001} {\bibfield  {journal} {\bibinfo  {journal}
  {Phys. Rev. Lett.}\ }\textbf {\bibinfo {volume} {91}},\ \bibinfo {pages}
  {107001} (\bibinfo {year} {2003})}\BibitemShut {NoStop}%
\bibitem [{\citenamefont {Itou}\ \emph {et~al.}(2008)\citenamefont {Itou},
  \citenamefont {Oyamada}, \citenamefont {Maegawa}, \citenamefont {Tamura},\
  and\ \citenamefont {Kato}}]{Itou2008}%
  \BibitemOpen
  \bibfield  {author} {\bibinfo {author} {\bibfnamefont {T.}~\bibnamefont
  {Itou}}, \bibinfo {author} {\bibfnamefont {A.}~\bibnamefont {Oyamada}},
  \bibinfo {author} {\bibfnamefont {S.}~\bibnamefont {Maegawa}}, \bibinfo
  {author} {\bibfnamefont {M.}~\bibnamefont {Tamura}}, \ and\ \bibinfo {author}
  {\bibfnamefont {R.}~\bibnamefont {Kato}},\ }\href {\doibase
  10.1103/PhysRevB.77.104413} {\bibfield  {journal} {\bibinfo  {journal} {Phys.
  Rev. B}\ }\textbf {\bibinfo {volume} {77}},\ \bibinfo {pages} {104413}
  (\bibinfo {year} {2008})}\BibitemShut {NoStop}%
\bibitem [{\citenamefont {Powell}\ and\ \citenamefont
  {McKenzie}(2011)}]{Powell2011}%
  \BibitemOpen
  \bibfield  {author} {\bibinfo {author} {\bibfnamefont {B.~J.}\ \bibnamefont
  {Powell}}\ and\ \bibinfo {author} {\bibfnamefont {R.~H.}\ \bibnamefont
  {McKenzie}},\ }\href {\doibase 10.1088/0034-4885/74/5/056501} {\bibfield
  {journal} {\bibinfo  {journal} {Rep. Prog. Phys.}\ }\textbf {\bibinfo
  {volume} {74}},\ \bibinfo {pages} {056501} (\bibinfo {year}
  {2011})}\BibitemShut {NoStop}%
\bibitem [{\citenamefont {Isono}\ \emph {et~al.}(2014)\citenamefont {Isono},
  \citenamefont {Kamo}, \citenamefont {Ueda}, \citenamefont {Takahashi},
  \citenamefont {Kimata}, \citenamefont {Tajima}, \citenamefont {Tsuchiya},
  \citenamefont {Terashima}, \citenamefont {Uji},\ and\ \citenamefont
  {Mori}}]{Isono2014}%
  \BibitemOpen
  \bibfield  {author} {\bibinfo {author} {\bibfnamefont {T.}~\bibnamefont
  {Isono}}, \bibinfo {author} {\bibfnamefont {H.}~\bibnamefont {Kamo}},
  \bibinfo {author} {\bibfnamefont {A.}~\bibnamefont {Ueda}}, \bibinfo {author}
  {\bibfnamefont {K.}~\bibnamefont {Takahashi}}, \bibinfo {author}
  {\bibfnamefont {M.}~\bibnamefont {Kimata}}, \bibinfo {author} {\bibfnamefont
  {H.}~\bibnamefont {Tajima}}, \bibinfo {author} {\bibfnamefont
  {S.}~\bibnamefont {Tsuchiya}}, \bibinfo {author} {\bibfnamefont
  {T.}~\bibnamefont {Terashima}}, \bibinfo {author} {\bibfnamefont
  {S.}~\bibnamefont {Uji}}, \ and\ \bibinfo {author} {\bibfnamefont
  {H.}~\bibnamefont {Mori}},\ }\href {\doibase 10.1103/PhysRevLett.112.177201}
  {\bibfield  {journal} {\bibinfo  {journal} {Phys. Rev. Lett.}\ }\textbf
  {\bibinfo {volume} {112}},\ \bibinfo {pages} {177201} (\bibinfo {year}
  {2014})}\BibitemShut {NoStop}%
\bibitem [{\citenamefont {Yamashita}\ \emph {et~al.}(2017)\citenamefont
  {Yamashita}, \citenamefont {Nakazawa}, \citenamefont {Ueda},\ and\
  \citenamefont {Mori}}]{Yamashita2017}%
  \BibitemOpen
  \bibfield  {author} {\bibinfo {author} {\bibfnamefont {S.}~\bibnamefont
  {Yamashita}}, \bibinfo {author} {\bibfnamefont {Y.}~\bibnamefont {Nakazawa}},
  \bibinfo {author} {\bibfnamefont {A.}~\bibnamefont {Ueda}}, \ and\ \bibinfo
  {author} {\bibfnamefont {H.}~\bibnamefont {Mori}},\ }\href {\doibase
  10.1103/PhysRevB.95.184425} {\bibfield  {journal} {\bibinfo  {journal} {Phys.
  Rev. B}\ }\textbf {\bibinfo {volume} {95}},\ \bibinfo {pages} {184425}
  (\bibinfo {year} {2017})}\BibitemShut {NoStop}%
\bibitem [{\citenamefont {Okamoto}\ \emph {et~al.}(2007)\citenamefont
  {Okamoto}, \citenamefont {Nohara}, \citenamefont {Aruga-Katori},\ and\
  \citenamefont {Takagi}}]{Okamoto2007}%
  \BibitemOpen
  \bibfield  {author} {\bibinfo {author} {\bibfnamefont {Y.}~\bibnamefont
  {Okamoto}}, \bibinfo {author} {\bibfnamefont {M.}~\bibnamefont {Nohara}},
  \bibinfo {author} {\bibfnamefont {H.}~\bibnamefont {Aruga-Katori}}, \ and\
  \bibinfo {author} {\bibfnamefont {H.}~\bibnamefont {Takagi}},\ }\href
  {\doibase 10.1103/PhysRevLett.99.137207} {\bibfield  {journal} {\bibinfo
  {journal} {Phys. Rev. Lett.}\ }\textbf {\bibinfo {volume} {99}},\ \bibinfo
  {pages} {137207} (\bibinfo {year} {2007})}\BibitemShut {NoStop}%
\bibitem [{\citenamefont {Jackeli}\ and\ \citenamefont
  {Khaliullin}(2009)}]{Jackeli2009}%
  \BibitemOpen
  \bibfield  {author} {\bibinfo {author} {\bibfnamefont {G.}~\bibnamefont
  {Jackeli}}\ and\ \bibinfo {author} {\bibfnamefont {G.}~\bibnamefont
  {Khaliullin}},\ }\href {\doibase 10.1103/PhysRevLett.102.017205} {\bibfield
  {journal} {\bibinfo  {journal} {Phys. Rev. Lett.}\ }\textbf {\bibinfo
  {volume} {102}},\ \bibinfo {pages} {017205} (\bibinfo {year}
  {2009})}\BibitemShut {NoStop}%
\bibitem [{\citenamefont {Chaloupka}\ \emph {et~al.}(2010)\citenamefont
  {Chaloupka}, \citenamefont {Jackeli},\ and\ \citenamefont
  {Khaliullin}}]{Chaloupka2010}%
  \BibitemOpen
  \bibfield  {author} {\bibinfo {author} {\bibfnamefont {J.}~\bibnamefont
  {Chaloupka}}, \bibinfo {author} {\bibfnamefont {G.}~\bibnamefont {Jackeli}},
  \ and\ \bibinfo {author} {\bibfnamefont {G.}~\bibnamefont {Khaliullin}},\
  }\href {\doibase 10.1103/PhysRevLett.105.027204} {\bibfield  {journal}
  {\bibinfo  {journal} {Phys. Rev. Lett.}\ }\textbf {\bibinfo {volume} {105}},\
  \bibinfo {pages} {027204} (\bibinfo {year} {2010})}\BibitemShut {NoStop}%
\bibitem [{\citenamefont {Singh}\ and\ \citenamefont
  {Gegenwart}(2010)}]{Singh2010}%
  \BibitemOpen
  \bibfield  {author} {\bibinfo {author} {\bibfnamefont {Y.}~\bibnamefont
  {Singh}}\ and\ \bibinfo {author} {\bibfnamefont {P.}~\bibnamefont
  {Gegenwart}},\ }\href {\doibase 10.1103/PhysRevB.82.064412} {\bibfield
  {journal} {\bibinfo  {journal} {Phys. Rev. B}\ }\textbf {\bibinfo {volume}
  {82}},\ \bibinfo {pages} {064412} (\bibinfo {year} {2010})}\BibitemShut
  {NoStop}%
\bibitem [{\citenamefont {Singh}\ \emph {et~al.}(2012)\citenamefont {Singh},
  \citenamefont {Manni}, \citenamefont {Reuther}, \citenamefont {Berlijn},
  \citenamefont {Thomale}, \citenamefont {Ku}, \citenamefont {Trebst},\ and\
  \citenamefont {Gegenwart}}]{Singh2012}%
  \BibitemOpen
  \bibfield  {author} {\bibinfo {author} {\bibfnamefont {Y.}~\bibnamefont
  {Singh}}, \bibinfo {author} {\bibfnamefont {S.}~\bibnamefont {Manni}},
  \bibinfo {author} {\bibfnamefont {J.}~\bibnamefont {Reuther}}, \bibinfo
  {author} {\bibfnamefont {T.}~\bibnamefont {Berlijn}}, \bibinfo {author}
  {\bibfnamefont {R.}~\bibnamefont {Thomale}}, \bibinfo {author} {\bibfnamefont
  {W.}~\bibnamefont {Ku}}, \bibinfo {author} {\bibfnamefont {S.}~\bibnamefont
  {Trebst}}, \ and\ \bibinfo {author} {\bibfnamefont {P.}~\bibnamefont
  {Gegenwart}},\ }\href {\doibase 10.1103/PhysRevLett.108.127203} {\bibfield
  {journal} {\bibinfo  {journal} {Phys. Rev. Lett.}\ }\textbf {\bibinfo
  {volume} {108}},\ \bibinfo {pages} {127203} (\bibinfo {year}
  {2012})}\BibitemShut {NoStop}%
\bibitem [{\citenamefont {Plumb}\ \emph {et~al.}(2014)\citenamefont {Plumb},
  \citenamefont {Clancy}, \citenamefont {Sandilands}, \citenamefont {Shankar},
  \citenamefont {Hu}, \citenamefont {Burch}, \citenamefont {Kee},\ and\
  \citenamefont {Kim}}]{Plumb2014}%
  \BibitemOpen
  \bibfield  {author} {\bibinfo {author} {\bibfnamefont {K.~W.}\ \bibnamefont
  {Plumb}}, \bibinfo {author} {\bibfnamefont {J.~P.}\ \bibnamefont {Clancy}},
  \bibinfo {author} {\bibfnamefont {L.~J.}\ \bibnamefont {Sandilands}},
  \bibinfo {author} {\bibfnamefont {V.~V.}\ \bibnamefont {Shankar}}, \bibinfo
  {author} {\bibfnamefont {Y.~F.}\ \bibnamefont {Hu}}, \bibinfo {author}
  {\bibfnamefont {K.~S.}\ \bibnamefont {Burch}}, \bibinfo {author}
  {\bibfnamefont {H.-Y.}\ \bibnamefont {Kee}}, \ and\ \bibinfo {author}
  {\bibfnamefont {Y.-J.}\ \bibnamefont {Kim}},\ }\href {\doibase
  10.1103/PhysRevB.90.041112} {\bibfield  {journal} {\bibinfo  {journal} {Phys.
  Rev. B}\ }\textbf {\bibinfo {volume} {90}},\ \bibinfo {pages} {041112}
  (\bibinfo {year} {2014})}\BibitemShut {NoStop}%
\bibitem [{\citenamefont {Sears}\ \emph {et~al.}(2015)\citenamefont {Sears},
  \citenamefont {Songvilay}, \citenamefont {Plumb}, \citenamefont {Clancy},
  \citenamefont {Qiu}, \citenamefont {Zhao}, \citenamefont {Parshall},\ and\
  \citenamefont {Kim}}]{Sears2015}%
  \BibitemOpen
  \bibfield  {author} {\bibinfo {author} {\bibfnamefont {J.~A.}\ \bibnamefont
  {Sears}}, \bibinfo {author} {\bibfnamefont {M.}~\bibnamefont {Songvilay}},
  \bibinfo {author} {\bibfnamefont {K.~W.}\ \bibnamefont {Plumb}}, \bibinfo
  {author} {\bibfnamefont {J.~P.}\ \bibnamefont {Clancy}}, \bibinfo {author}
  {\bibfnamefont {Y.}~\bibnamefont {Qiu}}, \bibinfo {author} {\bibfnamefont
  {Y.}~\bibnamefont {Zhao}}, \bibinfo {author} {\bibfnamefont {D.}~\bibnamefont
  {Parshall}}, \ and\ \bibinfo {author} {\bibfnamefont {Y.-J.}\ \bibnamefont
  {Kim}},\ }\href {\doibase 10.1103/PhysRevB.91.144420} {\bibfield  {journal}
  {\bibinfo  {journal} {Phys. Rev. B}\ }\textbf {\bibinfo {volume} {91}},\
  \bibinfo {pages} {144420} (\bibinfo {year} {2015})}\BibitemShut {NoStop}%
\bibitem [{\citenamefont {Rau}\ \emph {et~al.}(2016)\citenamefont {Rau},
  \citenamefont {Lee},\ and\ \citenamefont {Kee}}]{Rau2016}%
  \BibitemOpen
  \bibfield  {author} {\bibinfo {author} {\bibfnamefont {J.~G.}\ \bibnamefont
  {Rau}}, \bibinfo {author} {\bibfnamefont {E.~K.-H.}\ \bibnamefont {Lee}}, \
  and\ \bibinfo {author} {\bibfnamefont {H.-Y.}\ \bibnamefont {Kee}},\ }\href
  {\doibase 10.1146/annurev-conmatphys-031115-011319} {\bibfield  {journal}
  {\bibinfo  {journal} {Annu. Rev. Condens. Matter Phys.}\ }\textbf {\bibinfo
  {volume} {7}},\ \bibinfo {pages} {195} (\bibinfo {year} {2016})}\BibitemShut
  {NoStop}%
\bibitem [{\citenamefont {Trebst}(2017)}]{Trebst2017}%
  \BibitemOpen
  \bibfield  {author} {\bibinfo {author} {\bibfnamefont {S.}~\bibnamefont
  {Trebst}},\ }\href {https://arxiv.org/abs/1701.07056} {\bibfield  {journal}
  {\bibinfo  {journal} {arXiv:1701.07056}\ } (\bibinfo {year}
  {2017})}\BibitemShut {NoStop}%
\bibitem [{\citenamefont {Kitagawa}\ \emph {et~al.}(2018)\citenamefont
  {Kitagawa}, \citenamefont {Takayama}, \citenamefont {Matsumoto},
  \citenamefont {Kato}, \citenamefont {Takano}, \citenamefont {Kishimoto},
  \citenamefont {Bette}, \citenamefont {Dinnebier}, \citenamefont {Jackeli},\
  and\ \citenamefont {Takagi}}]{Kitagawa2018}%
  \BibitemOpen
  \bibfield  {author} {\bibinfo {author} {\bibfnamefont {K.}~\bibnamefont
  {Kitagawa}}, \bibinfo {author} {\bibfnamefont {T.}~\bibnamefont {Takayama}},
  \bibinfo {author} {\bibfnamefont {Y.}~\bibnamefont {Matsumoto}}, \bibinfo
  {author} {\bibfnamefont {A.}~\bibnamefont {Kato}}, \bibinfo {author}
  {\bibfnamefont {R.}~\bibnamefont {Takano}}, \bibinfo {author} {\bibfnamefont
  {Y.}~\bibnamefont {Kishimoto}}, \bibinfo {author} {\bibfnamefont
  {S.}~\bibnamefont {Bette}}, \bibinfo {author} {\bibfnamefont
  {R.}~\bibnamefont {Dinnebier}}, \bibinfo {author} {\bibfnamefont
  {G.}~\bibnamefont {Jackeli}}, \ and\ \bibinfo {author} {\bibfnamefont
  {H.}~\bibnamefont {Takagi}},\ }\href {https://doi.org/10.1038/nature25482}
  {\bibfield  {journal} {\bibinfo  {journal} {Nature}\ }\textbf {\bibinfo
  {volume} {554}},\ \bibinfo {pages} {341} (\bibinfo {year}
  {2018})}\BibitemShut {NoStop}%
\bibitem [{\citenamefont {Motome}\ and\ \citenamefont
  {Nasu}(2020)}]{Motome2019}%
  \BibitemOpen
  \bibfield  {author} {\bibinfo {author} {\bibfnamefont {Y.}~\bibnamefont
  {Motome}}\ and\ \bibinfo {author} {\bibfnamefont {J.}~\bibnamefont {Nasu}},\
  }\href {\doibase 10.7566/JPSJ.89.012002} {\bibfield  {journal} {\bibinfo
  {journal} {J. Phys. Soc. Jpn.}\ }\textbf {\bibinfo {volume} {89}},\ \bibinfo
  {pages} {012002} (\bibinfo {year} {2020})}\BibitemShut {NoStop}%
\bibitem [{\citenamefont {Knolle}\ \emph {et~al.}(2015)\citenamefont {Knolle},
  \citenamefont {Kovrizhin}, \citenamefont {Chalker},\ and\ \citenamefont
  {Moessner}}]{Knolle2015}%
  \BibitemOpen
  \bibfield  {author} {\bibinfo {author} {\bibfnamefont {J.}~\bibnamefont
  {Knolle}}, \bibinfo {author} {\bibfnamefont {D.~L.}\ \bibnamefont
  {Kovrizhin}}, \bibinfo {author} {\bibfnamefont {J.~T.}\ \bibnamefont
  {Chalker}}, \ and\ \bibinfo {author} {\bibfnamefont {R.}~\bibnamefont
  {Moessner}},\ }\href {\doibase 10.1103/PhysRevB.92.115127} {\bibfield
  {journal} {\bibinfo  {journal} {Phys. Rev. B}\ }\textbf {\bibinfo {volume}
  {92}},\ \bibinfo {pages} {115127} (\bibinfo {year} {2015})}\BibitemShut
  {NoStop}%
\bibitem [{\citenamefont {Knolle}\ \emph
  {et~al.}(2014{\natexlab{a}})\citenamefont {Knolle}, \citenamefont
  {Kovrizhin}, \citenamefont {Chalker},\ and\ \citenamefont
  {Moessner}}]{Knolle2014a}%
  \BibitemOpen
  \bibfield  {author} {\bibinfo {author} {\bibfnamefont {J.}~\bibnamefont
  {Knolle}}, \bibinfo {author} {\bibfnamefont {D.~L.}\ \bibnamefont
  {Kovrizhin}}, \bibinfo {author} {\bibfnamefont {J.~T.}\ \bibnamefont
  {Chalker}}, \ and\ \bibinfo {author} {\bibfnamefont {R.}~\bibnamefont
  {Moessner}},\ }\href {\doibase 10.1103/PhysRevLett.112.207203} {\bibfield
  {journal} {\bibinfo  {journal} {Phys. Rev. Lett.}\ }\textbf {\bibinfo
  {volume} {112}},\ \bibinfo {pages} {207203} (\bibinfo {year}
  {2014}{\natexlab{a}})}\BibitemShut {NoStop}%
\bibitem [{\citenamefont {Banerjee}\ \emph {et~al.}(2016)\citenamefont
  {Banerjee}, \citenamefont {Bridges}, \citenamefont {Yan}, \citenamefont
  {Aczel}, \citenamefont {Li}, \citenamefont {Stone}, \citenamefont {Granroth},
  \citenamefont {Lumsden}, \citenamefont {Yiu}, \citenamefont {Knolle},
  \citenamefont {Bhattacharjee}, \citenamefont {Kovrizhin}, \citenamefont
  {Moessner}, \citenamefont {Tennant}, \citenamefont {G.},\ and\ \citenamefont
  {Nagler}}]{Banerjee2016}%
  \BibitemOpen
  \bibfield  {author} {\bibinfo {author} {\bibfnamefont {A.}~\bibnamefont
  {Banerjee}}, \bibinfo {author} {\bibfnamefont {C.~A.}\ \bibnamefont
  {Bridges}}, \bibinfo {author} {\bibfnamefont {J.-Q.}\ \bibnamefont {Yan}},
  \bibinfo {author} {\bibfnamefont {A.~A.}\ \bibnamefont {Aczel}}, \bibinfo
  {author} {\bibfnamefont {L.}~\bibnamefont {Li}}, \bibinfo {author}
  {\bibfnamefont {M.~B.}\ \bibnamefont {Stone}}, \bibinfo {author}
  {\bibfnamefont {G.~E.}\ \bibnamefont {Granroth}}, \bibinfo {author}
  {\bibfnamefont {M.~D.}\ \bibnamefont {Lumsden}}, \bibinfo {author}
  {\bibfnamefont {Y.}~\bibnamefont {Yiu}}, \bibinfo {author} {\bibfnamefont
  {J.}~\bibnamefont {Knolle}}, \bibinfo {author} {\bibfnamefont
  {S.}~\bibnamefont {Bhattacharjee}}, \bibinfo {author} {\bibfnamefont {D.~L.}\
  \bibnamefont {Kovrizhin}}, \bibinfo {author} {\bibfnamefont {R.}~\bibnamefont
  {Moessner}}, \bibinfo {author} {\bibfnamefont {D.~A.}\ \bibnamefont
  {Tennant}}, \bibinfo {author} {\bibfnamefont {M.~D.}\ \bibnamefont {G.}}, \
  and\ \bibinfo {author} {\bibfnamefont {S.~E.}\ \bibnamefont {Nagler}},\
  }\href {https://doi.org/10.1038/nmat4604} {\bibfield  {journal} {\bibinfo
  {journal} {Nat. Mater.}\ } (\bibinfo {year} {2016})}\BibitemShut {NoStop}%
\bibitem [{\citenamefont {Banerjee}\ \emph {et~al.}(2017)\citenamefont
  {Banerjee}, \citenamefont {Yan}, \citenamefont {Knolle}, \citenamefont
  {Bridges}, \citenamefont {Stone}, \citenamefont {Lumsden}, \citenamefont
  {Mandrus}, \citenamefont {Tennant}, \citenamefont {Moessner},\ and\
  \citenamefont {Nagler}}]{Banerjee2017}%
  \BibitemOpen
  \bibfield  {author} {\bibinfo {author} {\bibfnamefont {A.}~\bibnamefont
  {Banerjee}}, \bibinfo {author} {\bibfnamefont {J.}~\bibnamefont {Yan}},
  \bibinfo {author} {\bibfnamefont {J.}~\bibnamefont {Knolle}}, \bibinfo
  {author} {\bibfnamefont {C.~A.}\ \bibnamefont {Bridges}}, \bibinfo {author}
  {\bibfnamefont {M.~B.}\ \bibnamefont {Stone}}, \bibinfo {author}
  {\bibfnamefont {M.~D.}\ \bibnamefont {Lumsden}}, \bibinfo {author}
  {\bibfnamefont {D.~G.}\ \bibnamefont {Mandrus}}, \bibinfo {author}
  {\bibfnamefont {D.~A.}\ \bibnamefont {Tennant}}, \bibinfo {author}
  {\bibfnamefont {R.}~\bibnamefont {Moessner}}, \ and\ \bibinfo {author}
  {\bibfnamefont {S.~E.}\ \bibnamefont {Nagler}},\ }\href
  {https://science.sciencemag.org/content/356/6342/1055} {\bibfield  {journal}
  {\bibinfo  {journal} {Science}\ }\textbf {\bibinfo {volume} {356}},\ \bibinfo
  {pages} {1055} (\bibinfo {year} {2017})}\BibitemShut {NoStop}%
\bibitem [{\citenamefont {Ko}\ \emph {et~al.}(2010)\citenamefont {Ko},
  \citenamefont {Liu}, \citenamefont {Ng},\ and\ \citenamefont {Lee}}]{Ko2010}%
  \BibitemOpen
  \bibfield  {author} {\bibinfo {author} {\bibfnamefont {W.-H.}\ \bibnamefont
  {Ko}}, \bibinfo {author} {\bibfnamefont {Z.-X.}\ \bibnamefont {Liu}},
  \bibinfo {author} {\bibfnamefont {T.-K.}\ \bibnamefont {Ng}}, \ and\ \bibinfo
  {author} {\bibfnamefont {P.~A.}\ \bibnamefont {Lee}},\ }\href {\doibase
  10.1103/PhysRevB.81.024414} {\bibfield  {journal} {\bibinfo  {journal} {Phys.
  Rev. B}\ }\textbf {\bibinfo {volume} {81}},\ \bibinfo {pages} {024414}
  (\bibinfo {year} {2010})}\BibitemShut {NoStop}%
\bibitem [{\citenamefont {Sandilands}\ \emph {et~al.}(2015)\citenamefont
  {Sandilands}, \citenamefont {Tian}, \citenamefont {Plumb}, \citenamefont
  {Kim},\ and\ \citenamefont {Burch}}]{Sandilands2015}%
  \BibitemOpen
  \bibfield  {author} {\bibinfo {author} {\bibfnamefont {L.~J.}\ \bibnamefont
  {Sandilands}}, \bibinfo {author} {\bibfnamefont {Y.}~\bibnamefont {Tian}},
  \bibinfo {author} {\bibfnamefont {K.~W.}\ \bibnamefont {Plumb}}, \bibinfo
  {author} {\bibfnamefont {Y.-J.}\ \bibnamefont {Kim}}, \ and\ \bibinfo
  {author} {\bibfnamefont {K.~S.}\ \bibnamefont {Burch}},\ }\href {\doibase
  10.1103/PhysRevLett.114.147201} {\bibfield  {journal} {\bibinfo  {journal}
  {Phys. Rev. Lett.}\ }\textbf {\bibinfo {volume} {114}},\ \bibinfo {pages}
  {147201} (\bibinfo {year} {2015})}\BibitemShut {NoStop}%
\bibitem [{\citenamefont {Knolle}\ \emph
  {et~al.}(2014{\natexlab{b}})\citenamefont {Knolle}, \citenamefont {Chern},
  \citenamefont {Kovrizhin}, \citenamefont {Moessner},\ and\ \citenamefont
  {Perkins}}]{Knolle2014b}%
  \BibitemOpen
  \bibfield  {author} {\bibinfo {author} {\bibfnamefont {J.}~\bibnamefont
  {Knolle}}, \bibinfo {author} {\bibfnamefont {G.-W.}\ \bibnamefont {Chern}},
  \bibinfo {author} {\bibfnamefont {D.~L.}\ \bibnamefont {Kovrizhin}}, \bibinfo
  {author} {\bibfnamefont {R.}~\bibnamefont {Moessner}}, \ and\ \bibinfo
  {author} {\bibfnamefont {N.~B.}\ \bibnamefont {Perkins}},\ }\href {\doibase
  10.1103/PhysRevLett.113.187201} {\bibfield  {journal} {\bibinfo  {journal}
  {Phys. Rev. Lett.}\ }\textbf {\bibinfo {volume} {113}},\ \bibinfo {pages}
  {187201} (\bibinfo {year} {2014}{\natexlab{b}})}\BibitemShut {NoStop}%
\bibitem [{\citenamefont {Nasu}\ \emph {et~al.}(2016)\citenamefont {Nasu},
  \citenamefont {Knolle}, \citenamefont {Kovrizhin}, \citenamefont {Motome},\
  and\ \citenamefont {Moessner}}]{Nasu2016}%
  \BibitemOpen
  \bibfield  {author} {\bibinfo {author} {\bibfnamefont {J.}~\bibnamefont
  {Nasu}}, \bibinfo {author} {\bibfnamefont {J.}~\bibnamefont {Knolle}},
  \bibinfo {author} {\bibfnamefont {D.~L.}\ \bibnamefont {Kovrizhin}}, \bibinfo
  {author} {\bibfnamefont {Y.}~\bibnamefont {Motome}}, \ and\ \bibinfo {author}
  {\bibfnamefont {R.}~\bibnamefont {Moessner}},\ }\href {\doibase
  http://dx.doi.org/10.1038/nphys3809} {\bibfield  {journal} {\bibinfo
  {journal} {Nat. Phys.}\ }\textbf {\bibinfo {volume} {12}},\ \bibinfo {pages}
  {912} (\bibinfo {year} {2016})}\BibitemShut {NoStop}%
\bibitem [{\citenamefont {Rousochatzakis}\ \emph {et~al.}(2019)\citenamefont
  {Rousochatzakis}, \citenamefont {Kourtis}, \citenamefont {Knolle},
  \citenamefont {Moessner},\ and\ \citenamefont
  {Perkins}}]{Rousochatzakis2019}%
  \BibitemOpen
  \bibfield  {author} {\bibinfo {author} {\bibfnamefont {I.}~\bibnamefont
  {Rousochatzakis}}, \bibinfo {author} {\bibfnamefont {S.}~\bibnamefont
  {Kourtis}}, \bibinfo {author} {\bibfnamefont {J.}~\bibnamefont {Knolle}},
  \bibinfo {author} {\bibfnamefont {R.}~\bibnamefont {Moessner}}, \ and\
  \bibinfo {author} {\bibfnamefont {N.~B.}\ \bibnamefont {Perkins}},\ }\href
  {\doibase 10.1103/PhysRevB.100.045117} {\bibfield  {journal} {\bibinfo
  {journal} {Phys. Rev. B}\ }\textbf {\bibinfo {volume} {100}},\ \bibinfo
  {pages} {045117} (\bibinfo {year} {2019})}\BibitemShut {NoStop}%
\bibitem [{\citenamefont {Sahasrabudhe}\ \emph {et~al.}(2020)\citenamefont
  {Sahasrabudhe}, \citenamefont {Kaib}, \citenamefont {Reschke}, \citenamefont
  {German}, \citenamefont {Koethe}, \citenamefont {Buhot}, \citenamefont
  {Kamenskyi}, \citenamefont {Hickey}, \citenamefont {Becker}, \citenamefont
  {Tsurkan}, \citenamefont {Loidl}, \citenamefont {Do}, \citenamefont {Choi},
  \citenamefont {Gr\"uninger}, \citenamefont {Winter}, \citenamefont {Wang},
  \citenamefont {Valent\'{\i}},\ and\ \citenamefont {van
  Loosdrecht}}]{Sahasrabudhe2020}%
  \BibitemOpen
  \bibfield  {author} {\bibinfo {author} {\bibfnamefont {A.}~\bibnamefont
  {Sahasrabudhe}}, \bibinfo {author} {\bibfnamefont {D.~A.~S.}\ \bibnamefont
  {Kaib}}, \bibinfo {author} {\bibfnamefont {S.}~\bibnamefont {Reschke}},
  \bibinfo {author} {\bibfnamefont {R.}~\bibnamefont {German}}, \bibinfo
  {author} {\bibfnamefont {T.~C.}\ \bibnamefont {Koethe}}, \bibinfo {author}
  {\bibfnamefont {J.}~\bibnamefont {Buhot}}, \bibinfo {author} {\bibfnamefont
  {D.}~\bibnamefont {Kamenskyi}}, \bibinfo {author} {\bibfnamefont
  {C.}~\bibnamefont {Hickey}}, \bibinfo {author} {\bibfnamefont
  {P.}~\bibnamefont {Becker}}, \bibinfo {author} {\bibfnamefont
  {V.}~\bibnamefont {Tsurkan}}, \bibinfo {author} {\bibfnamefont
  {A.}~\bibnamefont {Loidl}}, \bibinfo {author} {\bibfnamefont {S.~H.}\
  \bibnamefont {Do}}, \bibinfo {author} {\bibfnamefont {K.~Y.}\ \bibnamefont
  {Choi}}, \bibinfo {author} {\bibfnamefont {M.}~\bibnamefont {Gr\"uninger}},
  \bibinfo {author} {\bibfnamefont {S.~M.}\ \bibnamefont {Winter}}, \bibinfo
  {author} {\bibfnamefont {Z.}~\bibnamefont {Wang}}, \bibinfo {author}
  {\bibfnamefont {R.}~\bibnamefont {Valent\'{\i}}}, \ and\ \bibinfo {author}
  {\bibfnamefont {P.~H.~M.}\ \bibnamefont {van Loosdrecht}},\ }\href {\doibase
  10.1103/PhysRevB.101.140410} {\bibfield  {journal} {\bibinfo  {journal}
  {Phys. Rev. B}\ }\textbf {\bibinfo {volume} {101}},\ \bibinfo {pages}
  {140410} (\bibinfo {year} {2020})}\BibitemShut {NoStop}%
\bibitem [{\citenamefont {Wang}\ \emph {et~al.}(2020)\citenamefont {Wang},
  \citenamefont {Osterhoudt}, \citenamefont {Tian}, \citenamefont
  {Lampen-Kelley}, \citenamefont {Banerjee}, \citenamefont {Goldstein},
  \citenamefont {Yan}, \citenamefont {Knolle}, \citenamefont {Ji},
  \citenamefont {Cava}, \citenamefont {Nasu}, \citenamefont {Motome},
  \citenamefont {Nagler}, \citenamefont {Mandrus},\ and\ \citenamefont
  {Burch}}]{Yiping2020}%
  \BibitemOpen
  \bibfield  {author} {\bibinfo {author} {\bibfnamefont {Y.}~\bibnamefont
  {Wang}}, \bibinfo {author} {\bibfnamefont {G.~B.}\ \bibnamefont
  {Osterhoudt}}, \bibinfo {author} {\bibfnamefont {Y.}~\bibnamefont {Tian}},
  \bibinfo {author} {\bibfnamefont {P.}~\bibnamefont {Lampen-Kelley}}, \bibinfo
  {author} {\bibfnamefont {A.}~\bibnamefont {Banerjee}}, \bibinfo {author}
  {\bibfnamefont {T.}~\bibnamefont {Goldstein}}, \bibinfo {author}
  {\bibfnamefont {J.}~\bibnamefont {Yan}}, \bibinfo {author} {\bibfnamefont
  {J.}~\bibnamefont {Knolle}}, \bibinfo {author} {\bibfnamefont
  {H.}~\bibnamefont {Ji}}, \bibinfo {author} {\bibfnamefont {R.~J.}\
  \bibnamefont {Cava}}, \bibinfo {author} {\bibfnamefont {J.}~\bibnamefont
  {Nasu}}, \bibinfo {author} {\bibfnamefont {Y.}~\bibnamefont {Motome}},
  \bibinfo {author} {\bibfnamefont {S.~E.}\ \bibnamefont {Nagler}}, \bibinfo
  {author} {\bibfnamefont {D.}~\bibnamefont {Mandrus}}, \ and\ \bibinfo
  {author} {\bibfnamefont {K.~S.}\ \bibnamefont {Burch}},\ }\href {\doibase
  10.1038/s41535-020-0216-6} {\bibfield  {journal} {\bibinfo  {journal} {npj
  Quantum Mater.}\ }\textbf {\bibinfo {volume} {5}},\ \bibinfo {pages} {14}
  (\bibinfo {year} {2020})}\BibitemShut {NoStop}%
\bibitem [{\citenamefont {Wulferding}\ \emph {et~al.}(2020)\citenamefont
  {Wulferding}, \citenamefont {Choi}, \citenamefont {Do}, \citenamefont {Lee},
  \citenamefont {Lemmens}, \citenamefont {Faugeras},\ and\ \citenamefont
  {Gallais}}]{Dirk2020}%
  \BibitemOpen
  \bibfield  {author} {\bibinfo {author} {\bibfnamefont {D.}~\bibnamefont
  {Wulferding}}, \bibinfo {author} {\bibfnamefont {Y.}~\bibnamefont {Choi}},
  \bibinfo {author} {\bibfnamefont {S.-H.}\ \bibnamefont {Do}}, \bibinfo
  {author} {\bibfnamefont {C.~H.}\ \bibnamefont {Lee}}, \bibinfo {author}
  {\bibfnamefont {P.}~\bibnamefont {Lemmens}}, \bibinfo {author} {\bibfnamefont
  {C.}~\bibnamefont {Faugeras}}, \ and\ \bibinfo {author} {\bibfnamefont
  {K.-Y.}\ \bibnamefont {Gallais}, \bibfnamefont {Yann~andChoi}},\ }\href
  {\doibase 10.1038/s41467-020-15370-1} {\bibfield  {journal} {\bibinfo
  {journal} {Nat. Commun.}\ }\textbf {\bibinfo {volume} {11}},\ \bibinfo
  {pages} {1603} (\bibinfo {year} {2020})}\BibitemShut {NoStop}%
\bibitem [{\citenamefont {Hal\'asz}\ \emph {et~al.}(2016)\citenamefont
  {Hal\'asz}, \citenamefont {Perkins},\ and\ \citenamefont {van~den
  Brink}}]{Gabor2016}%
  \BibitemOpen
  \bibfield  {author} {\bibinfo {author} {\bibfnamefont {G.~B.}\ \bibnamefont
  {Hal\'asz}}, \bibinfo {author} {\bibfnamefont {N.~B.}\ \bibnamefont
  {Perkins}}, \ and\ \bibinfo {author} {\bibfnamefont {J.}~\bibnamefont
  {van~den Brink}},\ }\href {\doibase 10.1103/PhysRevLett.117.127203}
  {\bibfield  {journal} {\bibinfo  {journal} {Phys. Rev. Lett.}\ }\textbf
  {\bibinfo {volume} {117}},\ \bibinfo {pages} {127203} (\bibinfo {year}
  {2016})}\BibitemShut {NoStop}%
\bibitem [{\citenamefont {Hal\'asz}\ \emph {et~al.}(2017)\citenamefont
  {Hal\'asz}, \citenamefont {Perreault},\ and\ \citenamefont
  {Perkins}}]{Gabor2017}%
  \BibitemOpen
  \bibfield  {author} {\bibinfo {author} {\bibfnamefont {G.~B.}\ \bibnamefont
  {Hal\'asz}}, \bibinfo {author} {\bibfnamefont {B.}~\bibnamefont {Perreault}},
  \ and\ \bibinfo {author} {\bibfnamefont {N.~B.}\ \bibnamefont {Perkins}},\
  }\href {\doibase 10.1103/PhysRevLett.119.097202} {\bibfield  {journal}
  {\bibinfo  {journal} {Phys. Rev. Lett.}\ }\textbf {\bibinfo {volume} {119}},\
  \bibinfo {pages} {097202} (\bibinfo {year} {2017})}\BibitemShut {NoStop}%
\bibitem [{\citenamefont {Hal\'asz}\ \emph {et~al.}(2019)\citenamefont
  {Hal\'asz}, \citenamefont {Kourtis}, \citenamefont {Knolle},\ and\
  \citenamefont {Perkins}}]{Gabor2019}%
  \BibitemOpen
  \bibfield  {author} {\bibinfo {author} {\bibfnamefont {G.~B.}\ \bibnamefont
  {Hal\'asz}}, \bibinfo {author} {\bibfnamefont {S.}~\bibnamefont {Kourtis}},
  \bibinfo {author} {\bibfnamefont {J.}~\bibnamefont {Knolle}}, \ and\ \bibinfo
  {author} {\bibfnamefont {N.~B.}\ \bibnamefont {Perkins}},\ }\href {\doibase
  10.1103/PhysRevB.99.184417} {\bibfield  {journal} {\bibinfo  {journal} {Phys.
  Rev. B}\ }\textbf {\bibinfo {volume} {99}},\ \bibinfo {pages} {184417}
  (\bibinfo {year} {2019})}\BibitemShut {NoStop}%
\bibitem [{\citenamefont {Katsura}\ \emph {et~al.}(2010)\citenamefont
  {Katsura}, \citenamefont {Nagaosa},\ and\ \citenamefont {Lee}}]{Katsura2010}%
  \BibitemOpen
  \bibfield  {author} {\bibinfo {author} {\bibfnamefont {H.}~\bibnamefont
  {Katsura}}, \bibinfo {author} {\bibfnamefont {N.}~\bibnamefont {Nagaosa}}, \
  and\ \bibinfo {author} {\bibfnamefont {P.~A.}\ \bibnamefont {Lee}},\ }\href
  {\doibase 10.1103/PhysRevLett.104.066403} {\bibfield  {journal} {\bibinfo
  {journal} {Phys. Rev. Lett.}\ }\textbf {\bibinfo {volume} {104}},\ \bibinfo
  {pages} {066403} (\bibinfo {year} {2010})}\BibitemShut {NoStop}%
\bibitem [{\citenamefont {Nasu}\ \emph {et~al.}(2014)\citenamefont {Nasu},
  \citenamefont {Udagawa},\ and\ \citenamefont {Motome}}]{Nasu2014}%
  \BibitemOpen
  \bibfield  {author} {\bibinfo {author} {\bibfnamefont {J.}~\bibnamefont
  {Nasu}}, \bibinfo {author} {\bibfnamefont {M.}~\bibnamefont {Udagawa}}, \
  and\ \bibinfo {author} {\bibfnamefont {Y.}~\bibnamefont {Motome}},\ }\href
  {\doibase 10.1103/PhysRevLett.113.197205} {\bibfield  {journal} {\bibinfo
  {journal} {Phys. Rev. Lett.}\ }\textbf {\bibinfo {volume} {113}},\ \bibinfo
  {pages} {197205} (\bibinfo {year} {2014})}\BibitemShut {NoStop}%
\bibitem [{\citenamefont {Nasu}\ \emph {et~al.}(2015)\citenamefont {Nasu},
  \citenamefont {Udagawa},\ and\ \citenamefont {Motome}}]{Nasu2015}%
  \BibitemOpen
  \bibfield  {author} {\bibinfo {author} {\bibfnamefont {J.}~\bibnamefont
  {Nasu}}, \bibinfo {author} {\bibfnamefont {M.}~\bibnamefont {Udagawa}}, \
  and\ \bibinfo {author} {\bibfnamefont {Y.}~\bibnamefont {Motome}},\ }\href
  {\doibase 10.1103/PhysRevB.92.115122} {\bibfield  {journal} {\bibinfo
  {journal} {Phys. Rev. B}\ }\textbf {\bibinfo {volume} {92}},\ \bibinfo
  {pages} {115122} (\bibinfo {year} {2015})}\BibitemShut {NoStop}%
\bibitem [{\citenamefont {Hirobe}\ \emph {et~al.}(2017)\citenamefont {Hirobe},
  \citenamefont {Sato}, \citenamefont {Shiomi}, \citenamefont {Tanaka},\ and\
  \citenamefont {Saitoh}}]{Hirobe2017}%
  \BibitemOpen
  \bibfield  {author} {\bibinfo {author} {\bibfnamefont {D.}~\bibnamefont
  {Hirobe}}, \bibinfo {author} {\bibfnamefont {M.}~\bibnamefont {Sato}},
  \bibinfo {author} {\bibfnamefont {Y.}~\bibnamefont {Shiomi}}, \bibinfo
  {author} {\bibfnamefont {H.}~\bibnamefont {Tanaka}}, \ and\ \bibinfo {author}
  {\bibfnamefont {E.}~\bibnamefont {Saitoh}},\ }\href {\doibase
  10.1103/PhysRevB.95.241112} {\bibfield  {journal} {\bibinfo  {journal} {Phys.
  Rev. B}\ }\textbf {\bibinfo {volume} {95}},\ \bibinfo {pages} {241112}
  (\bibinfo {year} {2017})}\BibitemShut {NoStop}%
\bibitem [{\citenamefont {Kasahara}\ \emph
  {et~al.}(2018{\natexlab{a}})\citenamefont {Kasahara}, \citenamefont {Sugii},
  \citenamefont {Ohnishi}, \citenamefont {Shimozawa}, \citenamefont
  {Yamashita}, \citenamefont {Kurita}, \citenamefont {Tanaka}, \citenamefont
  {Nasu}, \citenamefont {Motome}, \citenamefont {Shibauchi},\ and\
  \citenamefont {Matsuda}}]{Kasaharaprl2018}%
  \BibitemOpen
  \bibfield  {author} {\bibinfo {author} {\bibfnamefont {Y.}~\bibnamefont
  {Kasahara}}, \bibinfo {author} {\bibfnamefont {K.}~\bibnamefont {Sugii}},
  \bibinfo {author} {\bibfnamefont {T.}~\bibnamefont {Ohnishi}}, \bibinfo
  {author} {\bibfnamefont {M.}~\bibnamefont {Shimozawa}}, \bibinfo {author}
  {\bibfnamefont {M.}~\bibnamefont {Yamashita}}, \bibinfo {author}
  {\bibfnamefont {N.}~\bibnamefont {Kurita}}, \bibinfo {author} {\bibfnamefont
  {H.}~\bibnamefont {Tanaka}}, \bibinfo {author} {\bibfnamefont
  {J.}~\bibnamefont {Nasu}}, \bibinfo {author} {\bibfnamefont {Y.}~\bibnamefont
  {Motome}}, \bibinfo {author} {\bibfnamefont {T.}~\bibnamefont {Shibauchi}}, \
  and\ \bibinfo {author} {\bibfnamefont {Y.}~\bibnamefont {Matsuda}},\ }\href
  {\doibase 10.1103/PhysRevLett.120.217205} {\bibfield  {journal} {\bibinfo
  {journal} {Phys. Rev. Lett.}\ }\textbf {\bibinfo {volume} {120}},\ \bibinfo
  {pages} {217205} (\bibinfo {year} {2018}{\natexlab{a}})}\BibitemShut
  {NoStop}%
\bibitem [{\citenamefont {Kasahara}\ \emph
  {et~al.}(2018{\natexlab{b}})\citenamefont {Kasahara}, \citenamefont
  {Ohnishi}, \citenamefont {Mizukami}, \citenamefont {Tanaka}, \citenamefont
  {Ma}, \citenamefont {Sugii}, \citenamefont {Kurita}, \citenamefont {Tanaka},
  \citenamefont {Nasu}, \citenamefont {Motome} \emph
  {et~al.}}]{kasahara2018majorana}%
  \BibitemOpen
  \bibfield  {author} {\bibinfo {author} {\bibfnamefont {Y.}~\bibnamefont
  {Kasahara}}, \bibinfo {author} {\bibfnamefont {T.}~\bibnamefont {Ohnishi}},
  \bibinfo {author} {\bibfnamefont {Y.}~\bibnamefont {Mizukami}}, \bibinfo
  {author} {\bibfnamefont {O.}~\bibnamefont {Tanaka}}, \bibinfo {author}
  {\bibfnamefont {S.}~\bibnamefont {Ma}}, \bibinfo {author} {\bibfnamefont
  {K.}~\bibnamefont {Sugii}}, \bibinfo {author} {\bibfnamefont
  {N.}~\bibnamefont {Kurita}}, \bibinfo {author} {\bibfnamefont
  {H.}~\bibnamefont {Tanaka}}, \bibinfo {author} {\bibfnamefont
  {J.}~\bibnamefont {Nasu}}, \bibinfo {author} {\bibfnamefont {Y.}~\bibnamefont
  {Motome}},  \emph {et~al.},\ }\href
  {https://doi.org/10.1038/s41586-018-0274-0} {\bibfield  {journal} {\bibinfo
  {journal} {Nature}\ }\textbf {\bibinfo {volume} {559}},\ \bibinfo {pages}
  {227} (\bibinfo {year} {2018}{\natexlab{b}})}\BibitemShut {NoStop}%
\bibitem [{\citenamefont {Nasu}\ \emph {et~al.}(2017)\citenamefont {Nasu},
  \citenamefont {Yoshitake},\ and\ \citenamefont {Motome}}]{Nasu2017}%
  \BibitemOpen
  \bibfield  {author} {\bibinfo {author} {\bibfnamefont {J.}~\bibnamefont
  {Nasu}}, \bibinfo {author} {\bibfnamefont {J.}~\bibnamefont {Yoshitake}}, \
  and\ \bibinfo {author} {\bibfnamefont {Y.}~\bibnamefont {Motome}},\ }\href
  {\doibase 10.1103/PhysRevLett.119.127204} {\bibfield  {journal} {\bibinfo
  {journal} {Phys. Rev. Lett.}\ }\textbf {\bibinfo {volume} {119}},\ \bibinfo
  {pages} {127204} (\bibinfo {year} {2017})}\BibitemShut {NoStop}%
\bibitem [{\citenamefont {Metavitsiadis}\ \emph {et~al.}(2017)\citenamefont
  {Metavitsiadis}, \citenamefont {Pidatella},\ and\ \citenamefont
  {Brenig}}]{Metavitsiadis2017}%
  \BibitemOpen
  \bibfield  {author} {\bibinfo {author} {\bibfnamefont {A.}~\bibnamefont
  {Metavitsiadis}}, \bibinfo {author} {\bibfnamefont {A.}~\bibnamefont
  {Pidatella}}, \ and\ \bibinfo {author} {\bibfnamefont {W.}~\bibnamefont
  {Brenig}},\ }\href {\doibase 10.1103/PhysRevB.96.205121} {\bibfield
  {journal} {\bibinfo  {journal} {Phys. Rev. B}\ }\textbf {\bibinfo {volume}
  {96}},\ \bibinfo {pages} {205121} (\bibinfo {year} {2017})}\BibitemShut
  {NoStop}%
\bibitem [{\citenamefont {Pidatella}\ \emph {et~al.}(2019)\citenamefont
  {Pidatella}, \citenamefont {Metavitsiadis},\ and\ \citenamefont
  {Brenig}}]{Pidatella2019}%
  \BibitemOpen
  \bibfield  {author} {\bibinfo {author} {\bibfnamefont {A.}~\bibnamefont
  {Pidatella}}, \bibinfo {author} {\bibfnamefont {A.}~\bibnamefont
  {Metavitsiadis}}, \ and\ \bibinfo {author} {\bibfnamefont {W.}~\bibnamefont
  {Brenig}},\ }\href {\doibase 10.1103/PhysRevB.99.075141} {\bibfield
  {journal} {\bibinfo  {journal} {Phys. Rev. B}\ }\textbf {\bibinfo {volume}
  {99}},\ \bibinfo {pages} {075141} (\bibinfo {year} {2019})}\BibitemShut
  {NoStop}%
\bibitem [{\citenamefont {Nasu}\ and\ \citenamefont
  {Motome}(2020)}]{Motome2020}%
  \BibitemOpen
  \bibfield  {author} {\bibinfo {author} {\bibfnamefont {J.}~\bibnamefont
  {Nasu}}\ and\ \bibinfo {author} {\bibfnamefont {Y.}~\bibnamefont {Motome}},\
  }\href {\doibase 10.1103/PhysRevB.102.054437} {\bibfield  {journal} {\bibinfo
   {journal} {Phys. Rev. B}\ }\textbf {\bibinfo {volume} {102}},\ \bibinfo
  {pages} {054437} (\bibinfo {year} {2020})}\BibitemShut {NoStop}%
\bibitem [{\citenamefont {Widmann}\ \emph {et~al.}(2019)\citenamefont
  {Widmann}, \citenamefont {Tsurkan}, \citenamefont {Prishchenko},
  \citenamefont {Mazurenko}, \citenamefont {Tsirlin},\ and\ \citenamefont
  {Loidl}}]{Widmann2019}%
  \BibitemOpen
  \bibfield  {author} {\bibinfo {author} {\bibfnamefont {S.}~\bibnamefont
  {Widmann}}, \bibinfo {author} {\bibfnamefont {V.}~\bibnamefont {Tsurkan}},
  \bibinfo {author} {\bibfnamefont {D.~A.}\ \bibnamefont {Prishchenko}},
  \bibinfo {author} {\bibfnamefont {V.~G.}\ \bibnamefont {Mazurenko}}, \bibinfo
  {author} {\bibfnamefont {A.~A.}\ \bibnamefont {Tsirlin}}, \ and\ \bibinfo
  {author} {\bibfnamefont {A.}~\bibnamefont {Loidl}},\ }\href {\doibase
  10.1103/PhysRevB.99.094415} {\bibfield  {journal} {\bibinfo  {journal} {Phys.
  Rev. B}\ }\textbf {\bibinfo {volume} {99}},\ \bibinfo {pages} {094415}
  (\bibinfo {year} {2019})}\BibitemShut {NoStop}%
\bibitem [{\citenamefont {Li}\ and\ \citenamefont {Chen}(2020)}]{LiChen2020}%
  \BibitemOpen
  \bibfield  {author} {\bibinfo {author} {\bibfnamefont {M.}~\bibnamefont
  {Li}}\ and\ \bibinfo {author} {\bibfnamefont {G.}~\bibnamefont {Chen}},\
  }\href {\doibase 10.1557/mrs.2020.124} {\bibfield  {journal} {\bibinfo
  {journal} {MRS Bulletin}\ }\textbf {\bibinfo {volume} {45}},\ \bibinfo
  {pages} {348} (\bibinfo {year} {2020})}\BibitemShut {NoStop}%
\bibitem [{\citenamefont {Feng}\ \emph {et~al.}(2020)\citenamefont {Feng},
  \citenamefont {Perkins},\ and\ \citenamefont {Burnell}}]{Feng2020}%
  \BibitemOpen
  \bibfield  {author} {\bibinfo {author} {\bibfnamefont {K.}~\bibnamefont
  {Feng}}, \bibinfo {author} {\bibfnamefont {N.~B.}\ \bibnamefont {Perkins}}, \
  and\ \bibinfo {author} {\bibfnamefont {F.~J.}\ \bibnamefont {Burnell}},\
  }\href {\doibase 10.1103/PhysRevB.102.224402} {\bibfield  {journal} {\bibinfo
   {journal} {Phys. Rev. B}\ }\textbf {\bibinfo {volume} {102}},\ \bibinfo
  {pages} {224402} (\bibinfo {year} {2020})}\BibitemShut {NoStop}%
\bibitem [{\citenamefont {Willans}\ \emph {et~al.}(2010)\citenamefont
  {Willans}, \citenamefont {Chalker},\ and\ \citenamefont
  {Moessner}}]{Willans2010}%
  \BibitemOpen
  \bibfield  {author} {\bibinfo {author} {\bibfnamefont {A.~J.}\ \bibnamefont
  {Willans}}, \bibinfo {author} {\bibfnamefont {J.~T.}\ \bibnamefont
  {Chalker}}, \ and\ \bibinfo {author} {\bibfnamefont {R.}~\bibnamefont
  {Moessner}},\ }\href {\doibase 10.1103/PhysRevLett.104.237203} {\bibfield
  {journal} {\bibinfo  {journal} {Phys. Rev. Lett.}\ }\textbf {\bibinfo
  {volume} {104}},\ \bibinfo {pages} {237203} (\bibinfo {year}
  {2010})}\BibitemShut {NoStop}%
\bibitem [{\citenamefont {Willans}\ \emph {et~al.}(2011)\citenamefont
  {Willans}, \citenamefont {Chalker},\ and\ \citenamefont
  {Moessner}}]{Willans2011}%
  \BibitemOpen
  \bibfield  {author} {\bibinfo {author} {\bibfnamefont {A.~J.}\ \bibnamefont
  {Willans}}, \bibinfo {author} {\bibfnamefont {J.~T.}\ \bibnamefont
  {Chalker}}, \ and\ \bibinfo {author} {\bibfnamefont {R.}~\bibnamefont
  {Moessner}},\ }\href {\doibase 10.1103/PhysRevB.84.115146} {\bibfield
  {journal} {\bibinfo  {journal} {Phys. Rev. B}\ }\textbf {\bibinfo {volume}
  {84}},\ \bibinfo {pages} {115146} (\bibinfo {year} {2011})}\BibitemShut
  {NoStop}%
\bibitem [{\citenamefont {Watanabe}\ \emph {et~al.}(2014)\citenamefont
  {Watanabe}, \citenamefont {Kawamura}, \citenamefont {Nakano},\ and\
  \citenamefont {Sakai}}]{Watanabe2014}%
  \BibitemOpen
  \bibfield  {author} {\bibinfo {author} {\bibfnamefont {K.}~\bibnamefont
  {Watanabe}}, \bibinfo {author} {\bibfnamefont {H.}~\bibnamefont {Kawamura}},
  \bibinfo {author} {\bibfnamefont {H.}~\bibnamefont {Nakano}}, \ and\ \bibinfo
  {author} {\bibfnamefont {T.}~\bibnamefont {Sakai}},\ }\href {\doibase
  10.7566/JPSJ.83.034714} {\bibfield  {journal} {\bibinfo  {journal} {J. Phys.
  Soc. Jpn.}\ }\textbf {\bibinfo {volume} {83}},\ \bibinfo {pages} {034714}
  (\bibinfo {year} {2014})}\BibitemShut {NoStop}%
\bibitem [{\citenamefont {Hal\'asz}\ \emph {et~al.}(2014)\citenamefont
  {Hal\'asz}, \citenamefont {Chalker},\ and\ \citenamefont
  {Moessner}}]{Gabor2014}%
  \BibitemOpen
  \bibfield  {author} {\bibinfo {author} {\bibfnamefont {G.~B.}\ \bibnamefont
  {Hal\'asz}}, \bibinfo {author} {\bibfnamefont {J.~T.}\ \bibnamefont
  {Chalker}}, \ and\ \bibinfo {author} {\bibfnamefont {R.}~\bibnamefont
  {Moessner}},\ }\href {\doibase 10.1103/PhysRevB.90.035145} {\bibfield
  {journal} {\bibinfo  {journal} {Phys. Rev. B}\ }\textbf {\bibinfo {volume}
  {90}},\ \bibinfo {pages} {035145} (\bibinfo {year} {2014})}\BibitemShut
  {NoStop}%
\bibitem [{\citenamefont {Zschocke}\ and\ \citenamefont
  {Vojta}(2015)}]{Zschocke2015}%
  \BibitemOpen
  \bibfield  {author} {\bibinfo {author} {\bibfnamefont {F.}~\bibnamefont
  {Zschocke}}\ and\ \bibinfo {author} {\bibfnamefont {M.}~\bibnamefont
  {Vojta}},\ }\href {\doibase 10.1103/PhysRevB.92.014403} {\bibfield  {journal}
  {\bibinfo  {journal} {Phys. Rev. B}\ }\textbf {\bibinfo {volume} {92}},\
  \bibinfo {pages} {014403} (\bibinfo {year} {2015})}\BibitemShut {NoStop}%
\bibitem [{\citenamefont {Sreejith}\ \emph {et~al.}(2016)\citenamefont
  {Sreejith}, \citenamefont {Bhattacharjee},\ and\ \citenamefont
  {Moessner}}]{Sreejith2016}%
  \BibitemOpen
  \bibfield  {author} {\bibinfo {author} {\bibfnamefont {G.~J.}\ \bibnamefont
  {Sreejith}}, \bibinfo {author} {\bibfnamefont {S.}~\bibnamefont
  {Bhattacharjee}}, \ and\ \bibinfo {author} {\bibfnamefont {R.}~\bibnamefont
  {Moessner}},\ }\href {\doibase 10.1103/PhysRevB.93.064433} {\bibfield
  {journal} {\bibinfo  {journal} {Phys. Rev. B}\ }\textbf {\bibinfo {volume}
  {93}},\ \bibinfo {pages} {064433} (\bibinfo {year} {2016})}\BibitemShut
  {NoStop}%
\bibitem [{\citenamefont {Savary}\ and\ \citenamefont
  {Balents}(2017{\natexlab{b}})}]{savary2017disorder}%
  \BibitemOpen
  \bibfield  {author} {\bibinfo {author} {\bibfnamefont {L.}~\bibnamefont
  {Savary}}\ and\ \bibinfo {author} {\bibfnamefont {L.}~\bibnamefont
  {Balents}},\ }\href {\doibase 10.1103/PhysRevLett.118.087203} {\bibfield
  {journal} {\bibinfo  {journal} {Phys. Rev. Lett.}\ }\textbf {\bibinfo
  {volume} {118}},\ \bibinfo {pages} {087203} (\bibinfo {year}
  {2017}{\natexlab{b}})}\BibitemShut {NoStop}%
\bibitem [{\citenamefont {Yamaguchi}\ \emph {et~al.}(2017)\citenamefont
  {Yamaguchi}, \citenamefont {Okada}, \citenamefont {Kono}, \citenamefont
  {Kittaka}, \citenamefont {Sakakibara}, \citenamefont {Okabe}, \citenamefont
  {Iwasaki},\ and\ \citenamefont {Hosokoshi}}]{Yamaguchi2017}%
  \BibitemOpen
  \bibfield  {author} {\bibinfo {author} {\bibfnamefont {H.}~\bibnamefont
  {Yamaguchi}}, \bibinfo {author} {\bibfnamefont {M.}~\bibnamefont {Okada}},
  \bibinfo {author} {\bibfnamefont {Y.}~\bibnamefont {Kono}}, \bibinfo {author}
  {\bibfnamefont {S.}~\bibnamefont {Kittaka}}, \bibinfo {author} {\bibfnamefont
  {T.}~\bibnamefont {Sakakibara}}, \bibinfo {author} {\bibfnamefont
  {T.}~\bibnamefont {Okabe}}, \bibinfo {author} {\bibfnamefont
  {Y.}~\bibnamefont {Iwasaki}}, \ and\ \bibinfo {author} {\bibfnamefont
  {Y.}~\bibnamefont {Hosokoshi}},\ }\href
  {https://doi.org/10.1038/s41598-017-16431-0} {\bibfield  {journal} {\bibinfo
  {journal} {Sci. Rep.}\ }\textbf {\bibinfo {volume} {7}},\ \bibinfo {pages}
  {16144} (\bibinfo {year} {2017})}\BibitemShut {NoStop}%
\bibitem [{\citenamefont {Kimchi}\ \emph {et~al.}(2018)\citenamefont {Kimchi},
  \citenamefont {Sheckelton}, \citenamefont {McQueen},\ and\ \citenamefont
  {Lee}}]{kimchi2018heat}%
  \BibitemOpen
  \bibfield  {author} {\bibinfo {author} {\bibfnamefont {I.}~\bibnamefont
  {Kimchi}}, \bibinfo {author} {\bibfnamefont {J.~P.}\ \bibnamefont
  {Sheckelton}}, \bibinfo {author} {\bibfnamefont {T.~M.}\ \bibnamefont
  {McQueen}}, \ and\ \bibinfo {author} {\bibfnamefont {P.~A.}\ \bibnamefont
  {Lee}},\ }\href {\doibase 10.1038/s41467-018-06800-2} {\bibfield  {journal}
  {\bibinfo  {journal} {Nat. Commun.}\ }\textbf {\bibinfo {volume} {9}},\
  \bibinfo {pages} {4367} (\bibinfo {year} {2018})}\BibitemShut {NoStop}%
\bibitem [{\citenamefont {Slagle}\ \emph {et~al.}(2018)\citenamefont {Slagle},
  \citenamefont {Choi}, \citenamefont {Chern},\ and\ \citenamefont
  {Kim}}]{Slagle2018theory}%
  \BibitemOpen
  \bibfield  {author} {\bibinfo {author} {\bibfnamefont {K.}~\bibnamefont
  {Slagle}}, \bibinfo {author} {\bibfnamefont {W.}~\bibnamefont {Choi}},
  \bibinfo {author} {\bibfnamefont {L.~E.}\ \bibnamefont {Chern}}, \ and\
  \bibinfo {author} {\bibfnamefont {Y.~B.}\ \bibnamefont {Kim}},\ }\href
  {\doibase 10.1103/PhysRevB.97.115159} {\bibfield  {journal} {\bibinfo
  {journal} {Phys. Rev. B}\ }\textbf {\bibinfo {volume} {97}},\ \bibinfo
  {pages} {115159} (\bibinfo {year} {2018})}\BibitemShut {NoStop}%
\bibitem [{\citenamefont {Li}\ \emph {et~al.}(2018)\citenamefont {Li},
  \citenamefont {Winter},\ and\ \citenamefont {Valent\'{\i}}}]{li2018role}%
  \BibitemOpen
  \bibfield  {author} {\bibinfo {author} {\bibfnamefont {Y.}~\bibnamefont
  {Li}}, \bibinfo {author} {\bibfnamefont {S.~M.}\ \bibnamefont {Winter}}, \
  and\ \bibinfo {author} {\bibfnamefont {R.}~\bibnamefont {Valent\'{\i}}},\
  }\href {\doibase 10.1103/PhysRevLett.121.247202} {\bibfield  {journal}
  {\bibinfo  {journal} {Phys. Rev. Lett.}\ }\textbf {\bibinfo {volume} {121}},\
  \bibinfo {pages} {247202} (\bibinfo {year} {2018})}\BibitemShut {NoStop}%
\bibitem [{\citenamefont {Knolle}\ \emph {et~al.}(2019)\citenamefont {Knolle},
  \citenamefont {Moessner},\ and\ \citenamefont {Perkins}}]{Knolle2019}%
  \BibitemOpen
  \bibfield  {author} {\bibinfo {author} {\bibfnamefont {J.}~\bibnamefont
  {Knolle}}, \bibinfo {author} {\bibfnamefont {R.}~\bibnamefont {Moessner}}, \
  and\ \bibinfo {author} {\bibfnamefont {N.~B.}\ \bibnamefont {Perkins}},\
  }\href {\doibase 10.1103/PhysRevLett.122.047202} {\bibfield  {journal}
  {\bibinfo  {journal} {Phys. Rev. Lett.}\ }\textbf {\bibinfo {volume} {122}},\
  \bibinfo {pages} {047202} (\bibinfo {year} {2019})}\BibitemShut {NoStop}%
\bibitem [{\citenamefont {Takahashi}\ \emph {et~al.}(2019)\citenamefont
  {Takahashi}, \citenamefont {Wang}, \citenamefont {Arsenault}, \citenamefont
  {Imai}, \citenamefont {Abramchuk}, \citenamefont {Tafti},\ and\ \citenamefont
  {Singer}}]{Takahashi2019}%
  \BibitemOpen
  \bibfield  {author} {\bibinfo {author} {\bibfnamefont {S.~K.}\ \bibnamefont
  {Takahashi}}, \bibinfo {author} {\bibfnamefont {J.}~\bibnamefont {Wang}},
  \bibinfo {author} {\bibfnamefont {A.}~\bibnamefont {Arsenault}}, \bibinfo
  {author} {\bibfnamefont {T.}~\bibnamefont {Imai}}, \bibinfo {author}
  {\bibfnamefont {M.}~\bibnamefont {Abramchuk}}, \bibinfo {author}
  {\bibfnamefont {F.}~\bibnamefont {Tafti}}, \ and\ \bibinfo {author}
  {\bibfnamefont {P.~M.}\ \bibnamefont {Singer}},\ }\href {\doibase
  10.1103/PhysRevX.9.031047} {\bibfield  {journal} {\bibinfo  {journal} {Phys.
  Rev. X}\ }\textbf {\bibinfo {volume} {9}},\ \bibinfo {pages} {031047}
  (\bibinfo {year} {2019})}\BibitemShut {NoStop}%
\bibitem [{\citenamefont {Do}\ \emph {et~al.}(2020)\citenamefont {Do},
  \citenamefont {Lee}, \citenamefont {Kihara}, \citenamefont {Choi},
  \citenamefont {Yoon}, \citenamefont {Kim}, \citenamefont {Cheong},
  \citenamefont {Chen}, \citenamefont {Chou}, \citenamefont {Nojiri},\ and\
  \citenamefont {Choi}}]{Do2020}%
  \BibitemOpen
  \bibfield  {author} {\bibinfo {author} {\bibfnamefont {S.-H.}\ \bibnamefont
  {Do}}, \bibinfo {author} {\bibfnamefont {C.~H.}\ \bibnamefont {Lee}},
  \bibinfo {author} {\bibfnamefont {T.}~\bibnamefont {Kihara}}, \bibinfo
  {author} {\bibfnamefont {Y.~S.}\ \bibnamefont {Choi}}, \bibinfo {author}
  {\bibfnamefont {S.}~\bibnamefont {Yoon}}, \bibinfo {author} {\bibfnamefont
  {K.}~\bibnamefont {Kim}}, \bibinfo {author} {\bibfnamefont {H.}~\bibnamefont
  {Cheong}}, \bibinfo {author} {\bibfnamefont {W.-T.}\ \bibnamefont {Chen}},
  \bibinfo {author} {\bibfnamefont {F.}~\bibnamefont {Chou}}, \bibinfo {author}
  {\bibfnamefont {H.}~\bibnamefont {Nojiri}}, \ and\ \bibinfo {author}
  {\bibfnamefont {K.-Y.}\ \bibnamefont {Choi}},\ }\href {\doibase
  10.1103/PhysRevLett.124.047204} {\bibfield  {journal} {\bibinfo  {journal}
  {Phys. Rev. Lett.}\ }\textbf {\bibinfo {volume} {124}},\ \bibinfo {pages}
  {047204} (\bibinfo {year} {2020})}\BibitemShut {NoStop}%
\bibitem [{\citenamefont {Murayama}\ \emph {et~al.}(2020)\citenamefont
  {Murayama}, \citenamefont {Sato}, \citenamefont {Taniguchi}, \citenamefont
  {Kurihara}, \citenamefont {Xing}, \citenamefont {Huang}, \citenamefont
  {Kasahara}, \citenamefont {Kasahara}, \citenamefont {Kimchi}, \citenamefont
  {Yoshida}, \citenamefont {Iwasa}, \citenamefont {Mizukami}, \citenamefont
  {Shibauchi}, \citenamefont {Konczykowski},\ and\ \citenamefont
  {Matsuda}}]{Murayama2020}%
  \BibitemOpen
  \bibfield  {author} {\bibinfo {author} {\bibfnamefont {H.}~\bibnamefont
  {Murayama}}, \bibinfo {author} {\bibfnamefont {Y.}~\bibnamefont {Sato}},
  \bibinfo {author} {\bibfnamefont {T.}~\bibnamefont {Taniguchi}}, \bibinfo
  {author} {\bibfnamefont {R.}~\bibnamefont {Kurihara}}, \bibinfo {author}
  {\bibfnamefont {X.~Z.}\ \bibnamefont {Xing}}, \bibinfo {author}
  {\bibfnamefont {W.}~\bibnamefont {Huang}}, \bibinfo {author} {\bibfnamefont
  {S.}~\bibnamefont {Kasahara}}, \bibinfo {author} {\bibfnamefont
  {Y.}~\bibnamefont {Kasahara}}, \bibinfo {author} {\bibfnamefont
  {I.}~\bibnamefont {Kimchi}}, \bibinfo {author} {\bibfnamefont
  {M.}~\bibnamefont {Yoshida}}, \bibinfo {author} {\bibfnamefont
  {Y.}~\bibnamefont {Iwasa}}, \bibinfo {author} {\bibfnamefont
  {Y.}~\bibnamefont {Mizukami}}, \bibinfo {author} {\bibfnamefont
  {T.}~\bibnamefont {Shibauchi}}, \bibinfo {author} {\bibfnamefont
  {M.}~\bibnamefont {Konczykowski}}, \ and\ \bibinfo {author} {\bibfnamefont
  {Y.}~\bibnamefont {Matsuda}},\ }\href {\doibase
  10.1103/PhysRevResearch.2.013099} {\bibfield  {journal} {\bibinfo  {journal}
  {Phys. Rev. Research}\ }\textbf {\bibinfo {volume} {2}},\ \bibinfo {pages}
  {013099} (\bibinfo {year} {2020})}\BibitemShut {NoStop}%
\bibitem [{\citenamefont {Yamada}(2020)}]{Yamada2020}%
  \BibitemOpen
  \bibfield  {author} {\bibinfo {author} {\bibfnamefont {M.~G.}\ \bibnamefont
  {Yamada}},\ }\href {\doibase 10.1038/s41535-020-00285-3} {\bibfield
  {journal} {\bibinfo  {journal} {npj Quantum Mater.}\ }\textbf {\bibinfo
  {volume} {5}},\ \bibinfo {pages} {82} (\bibinfo {year} {2020})}\BibitemShut
  {NoStop}%
\bibitem [{\citenamefont {Kao}\ \emph {et~al.}(2021)\citenamefont {Kao},
  \citenamefont {Knolle}, \citenamefont {Hal\'asz}, \citenamefont {Moessner},\
  and\ \citenamefont {Perkins}}]{Kao2020}%
  \BibitemOpen
  \bibfield  {author} {\bibinfo {author} {\bibfnamefont {W.-H.}\ \bibnamefont
  {Kao}}, \bibinfo {author} {\bibfnamefont {J.}~\bibnamefont {Knolle}},
  \bibinfo {author} {\bibfnamefont {G.~B.}\ \bibnamefont {Hal\'asz}}, \bibinfo
  {author} {\bibfnamefont {R.}~\bibnamefont {Moessner}}, \ and\ \bibinfo
  {author} {\bibfnamefont {N.~B.}\ \bibnamefont {Perkins}},\ }\href {\doibase
  10.1103/PhysRevX.11.011034} {\bibfield  {journal} {\bibinfo  {journal} {Phys.
  Rev. X}\ }\textbf {\bibinfo {volume} {11}},\ \bibinfo {pages} {011034}
  (\bibinfo {year} {2021})}\BibitemShut {NoStop}%
\bibitem [{\citenamefont {Hermanns}\ \emph {et~al.}(2018)\citenamefont
  {Hermanns}, \citenamefont {Kimchi},\ and\ \citenamefont
  {Knolle}}]{hermanns2018}%
  \BibitemOpen
  \bibfield  {author} {\bibinfo {author} {\bibfnamefont {M.}~\bibnamefont
  {Hermanns}}, \bibinfo {author} {\bibfnamefont {I.}~\bibnamefont {Kimchi}}, \
  and\ \bibinfo {author} {\bibfnamefont {J.}~\bibnamefont {Knolle}},\ }\href
  {\doibase 10.1146/annurev-conmatphys-033117-053934} {\bibfield  {journal}
  {\bibinfo  {journal} {Annu. Rev. Condens. Matter Phys.}\ }\textbf {\bibinfo
  {volume} {9}},\ \bibinfo {pages} {17} (\bibinfo {year} {2018})}\BibitemShut
  {NoStop}%
\bibitem [{\citenamefont {Majumder}\ \emph {et~al.}(2015)\citenamefont
  {Majumder}, \citenamefont {Schmidt}, \citenamefont {Rosner}, \citenamefont
  {Tsirlin}, \citenamefont {Yasuoka},\ and\ \citenamefont
  {Baenitz}}]{Majumder2015}%
  \BibitemOpen
  \bibfield  {author} {\bibinfo {author} {\bibfnamefont {M.}~\bibnamefont
  {Majumder}}, \bibinfo {author} {\bibfnamefont {M.}~\bibnamefont {Schmidt}},
  \bibinfo {author} {\bibfnamefont {H.}~\bibnamefont {Rosner}}, \bibinfo
  {author} {\bibfnamefont {A.~A.}\ \bibnamefont {Tsirlin}}, \bibinfo {author}
  {\bibfnamefont {H.}~\bibnamefont {Yasuoka}}, \ and\ \bibinfo {author}
  {\bibfnamefont {M.}~\bibnamefont {Baenitz}},\ }\href {\doibase
  10.1103/PhysRevB.91.180401} {\bibfield  {journal} {\bibinfo  {journal} {Phys.
  Rev. B}\ }\textbf {\bibinfo {volume} {91}},\ \bibinfo {pages} {180401}
  (\bibinfo {year} {2015})}\BibitemShut {NoStop}%
\bibitem [{\citenamefont {Johnson}\ \emph {et~al.}(2015)\citenamefont
  {Johnson}, \citenamefont {Williams}, \citenamefont {Haghighirad},
  \citenamefont {Singleton}, \citenamefont {Zapf}, \citenamefont {Manuel},
  \citenamefont {Mazin}, \citenamefont {Li}, \citenamefont {Jeschke},
  \citenamefont {Valent\'{\i}},\ and\ \citenamefont {Coldea}}]{Johnson2015}%
  \BibitemOpen
  \bibfield  {author} {\bibinfo {author} {\bibfnamefont {R.~D.}\ \bibnamefont
  {Johnson}}, \bibinfo {author} {\bibfnamefont {S.~C.}\ \bibnamefont
  {Williams}}, \bibinfo {author} {\bibfnamefont {A.~A.}\ \bibnamefont
  {Haghighirad}}, \bibinfo {author} {\bibfnamefont {J.}~\bibnamefont
  {Singleton}}, \bibinfo {author} {\bibfnamefont {V.}~\bibnamefont {Zapf}},
  \bibinfo {author} {\bibfnamefont {P.}~\bibnamefont {Manuel}}, \bibinfo
  {author} {\bibfnamefont {I.~I.}\ \bibnamefont {Mazin}}, \bibinfo {author}
  {\bibfnamefont {Y.}~\bibnamefont {Li}}, \bibinfo {author} {\bibfnamefont
  {H.~O.}\ \bibnamefont {Jeschke}}, \bibinfo {author} {\bibfnamefont
  {R.}~\bibnamefont {Valent\'{\i}}}, \ and\ \bibinfo {author} {\bibfnamefont
  {R.}~\bibnamefont {Coldea}},\ }\href {\doibase 10.1103/PhysRevB.92.235119}
  {\bibfield  {journal} {\bibinfo  {journal} {Phys. Rev. B}\ }\textbf {\bibinfo
  {volume} {92}},\ \bibinfo {pages} {235119} (\bibinfo {year}
  {2015})}\BibitemShut {NoStop}%
\bibitem [{\citenamefont {Bahrami}\ \emph {et~al.}(2019)\citenamefont
  {Bahrami}, \citenamefont {Lafargue-Dit-Hauret}, \citenamefont {Lebedev},
  \citenamefont {Movshovich}, \citenamefont {Yang}, \citenamefont {Broido},
  \citenamefont {Rocquefelte},\ and\ \citenamefont {Tafti}}]{Tafti2019}%
  \BibitemOpen
  \bibfield  {author} {\bibinfo {author} {\bibfnamefont {F.}~\bibnamefont
  {Bahrami}}, \bibinfo {author} {\bibfnamefont {W.}~\bibnamefont
  {Lafargue-Dit-Hauret}}, \bibinfo {author} {\bibfnamefont {O.~I.}\
  \bibnamefont {Lebedev}}, \bibinfo {author} {\bibfnamefont {R.}~\bibnamefont
  {Movshovich}}, \bibinfo {author} {\bibfnamefont {H.-Y.}\ \bibnamefont
  {Yang}}, \bibinfo {author} {\bibfnamefont {D.}~\bibnamefont {Broido}},
  \bibinfo {author} {\bibfnamefont {X.}~\bibnamefont {Rocquefelte}}, \ and\
  \bibinfo {author} {\bibfnamefont {F.}~\bibnamefont {Tafti}},\ }\href
  {\doibase 10.1103/PhysRevLett.123.237203} {\bibfield  {journal} {\bibinfo
  {journal} {Phys. Rev. Lett.}\ }\textbf {\bibinfo {volume} {123}},\ \bibinfo
  {pages} {237203} (\bibinfo {year} {2019})}\BibitemShut {NoStop}%
\bibitem [{\citenamefont {Anderson}(1958)}]{Anderson1958}%
  \BibitemOpen
  \bibfield  {author} {\bibinfo {author} {\bibfnamefont {P.~W.}\ \bibnamefont
  {Anderson}},\ }\href {\doibase 10.1103/PhysRev.109.1492} {\bibfield
  {journal} {\bibinfo  {journal} {Phys. Rev.}\ }\textbf {\bibinfo {volume}
  {109}},\ \bibinfo {pages} {1492} (\bibinfo {year} {1958})}\BibitemShut
  {NoStop}%
\bibitem [{\citenamefont {Mott}(1967)}]{Mott1967}%
  \BibitemOpen
  \bibfield  {author} {\bibinfo {author} {\bibfnamefont {N.}~\bibnamefont
  {Mott}},\ }\href {\doibase 10.1080/00018736700101265} {\bibfield  {journal}
  {\bibinfo  {journal} {Adv. Phys.}\ }\textbf {\bibinfo {volume} {16}},\
  \bibinfo {pages} {49} (\bibinfo {year} {1967})}\BibitemShut {NoStop}%
\bibitem [{\citenamefont {Abrahams}\ \emph {et~al.}(1979)\citenamefont
  {Abrahams}, \citenamefont {Anderson}, \citenamefont {Licciardello},\ and\
  \citenamefont {Ramakrishnan}}]{Abrahams1979}%
  \BibitemOpen
  \bibfield  {author} {\bibinfo {author} {\bibfnamefont {E.}~\bibnamefont
  {Abrahams}}, \bibinfo {author} {\bibfnamefont {P.~W.}\ \bibnamefont
  {Anderson}}, \bibinfo {author} {\bibfnamefont {D.~C.}\ \bibnamefont
  {Licciardello}}, \ and\ \bibinfo {author} {\bibfnamefont {T.~V.}\
  \bibnamefont {Ramakrishnan}},\ }\href {\doibase 10.1103/PhysRevLett.42.673}
  {\bibfield  {journal} {\bibinfo  {journal} {Phys. Rev. Lett.}\ }\textbf
  {\bibinfo {volume} {42}},\ \bibinfo {pages} {673} (\bibinfo {year}
  {1979})}\BibitemShut {NoStop}%
\bibitem [{\citenamefont {Lee}\ and\ \citenamefont
  {Ramakrishnan}(1985)}]{Lee1985}%
  \BibitemOpen
  \bibfield  {author} {\bibinfo {author} {\bibfnamefont {P.~A.}\ \bibnamefont
  {Lee}}\ and\ \bibinfo {author} {\bibfnamefont {T.~V.}\ \bibnamefont
  {Ramakrishnan}},\ }\href {\doibase 10.1103/RevModPhys.57.287} {\bibfield
  {journal} {\bibinfo  {journal} {Rev. Mod. Phys.}\ }\textbf {\bibinfo {volume}
  {57}},\ \bibinfo {pages} {287} (\bibinfo {year} {1985})}\BibitemShut
  {NoStop}%
\bibitem [{\citenamefont {Shklovskii}\ and\ \citenamefont
  {Efros}(1984)}]{Boris1984}%
  \BibitemOpen
  \bibfield  {author} {\bibinfo {author} {\bibfnamefont {B.~I.}\ \bibnamefont
  {Shklovskii}}\ and\ \bibinfo {author} {\bibfnamefont {A.~L.}\ \bibnamefont
  {Efros}},\ }\href@noop {} {\emph {\bibinfo {title} {Electronic Properties of
  Doped Semiconductors}}}\ (\bibinfo  {publisher} {Springer},\ \bibinfo {year}
  {1984})\BibitemShut {NoStop}%
\bibitem [{\citenamefont {Kramer}\ and\ \citenamefont
  {MacKinnon}(1993)}]{Kramer1993}%
  \BibitemOpen
  \bibfield  {author} {\bibinfo {author} {\bibfnamefont {B.}~\bibnamefont
  {Kramer}}\ and\ \bibinfo {author} {\bibfnamefont {A.}~\bibnamefont
  {MacKinnon}},\ }\href {\doibase 10.1088/0034-4885/56/12/001} {\bibfield
  {journal} {\bibinfo  {journal} {Rep. Prog. Phys.}\ }\textbf {\bibinfo
  {volume} {56}},\ \bibinfo {pages} {1469} (\bibinfo {year}
  {1993})}\BibitemShut {NoStop}%
\bibitem [{\citenamefont {Abrahams}(2010)}]{50Anderson}%
  \BibitemOpen
  \bibfield  {author} {\bibinfo {author} {\bibfnamefont {E.}~\bibnamefont
  {Abrahams}},\ }\href@noop {} {\emph {\bibinfo {title} {50 Years of Anderson
  Localization}}}\ (\bibinfo  {publisher} {World Scientific Publishing Co},\
  \bibinfo {year} {2010})\BibitemShut {NoStop}%
\bibitem [{\citenamefont {Evers}\ and\ \citenamefont
  {Mirlin}(2008)}]{Evers2008}%
  \BibitemOpen
  \bibfield  {author} {\bibinfo {author} {\bibfnamefont {F.}~\bibnamefont
  {Evers}}\ and\ \bibinfo {author} {\bibfnamefont {A.~D.}\ \bibnamefont
  {Mirlin}},\ }\href {\doibase 10.1103/RevModPhys.80.1355} {\bibfield
  {journal} {\bibinfo  {journal} {Rev. Mod. Phys.}\ }\textbf {\bibinfo {volume}
  {80}},\ \bibinfo {pages} {1355} (\bibinfo {year} {2008})}\BibitemShut
  {NoStop}%
\bibitem [{\citenamefont {Baskaran}\ \emph {et~al.}(2007)\citenamefont
  {Baskaran}, \citenamefont {Mandal},\ and\ \citenamefont
  {Shankar}}]{Baskaran2007}%
  \BibitemOpen
  \bibfield  {author} {\bibinfo {author} {\bibfnamefont {G.}~\bibnamefont
  {Baskaran}}, \bibinfo {author} {\bibfnamefont {S.}~\bibnamefont {Mandal}}, \
  and\ \bibinfo {author} {\bibfnamefont {R.}~\bibnamefont {Shankar}},\ }\href
  {\doibase 10.1103/PhysRevLett.98.247201} {\bibfield  {journal} {\bibinfo
  {journal} {Phys. Rev. Lett.}\ }\textbf {\bibinfo {volume} {98}},\ \bibinfo
  {pages} {247201} (\bibinfo {year} {2007})}\BibitemShut {NoStop}%
\bibitem [{\citenamefont {Pereira}\ \emph {et~al.}(2008)\citenamefont
  {Pereira}, \citenamefont {Lopes~dos Santos},\ and\ \citenamefont
  {Castro~Neto}}]{Pereira2008}%
  \BibitemOpen
  \bibfield  {author} {\bibinfo {author} {\bibfnamefont {V.~M.}\ \bibnamefont
  {Pereira}}, \bibinfo {author} {\bibfnamefont {J.~M.~B.}\ \bibnamefont
  {Lopes~dos Santos}}, \ and\ \bibinfo {author} {\bibfnamefont {A.~H.}\
  \bibnamefont {Castro~Neto}},\ }\href {\doibase 10.1103/PhysRevB.77.115109}
  {\bibfield  {journal} {\bibinfo  {journal} {Phys. Rev. B}\ }\textbf {\bibinfo
  {volume} {77}},\ \bibinfo {pages} {115109} (\bibinfo {year}
  {2008})}\BibitemShut {NoStop}%
\bibitem [{\citenamefont {Lifshitz}(1965)}]{Lifshitz1965}%
  \BibitemOpen
  \bibfield  {author} {\bibinfo {author} {\bibfnamefont {I.~M.}\ \bibnamefont
  {Lifshitz}},\ }\href {\doibase 10.1070/pu1965v007n04abeh003634} {\bibfield
  {journal} {\bibinfo  {journal} {Soviet Physics Uspekhi}\ }\textbf {\bibinfo
  {volume} {7}},\ \bibinfo {pages} {549} (\bibinfo {year} {1965})}\BibitemShut
  {NoStop}%
\bibitem [{\citenamefont {Feng}\ \emph {et~al.}(2021)\citenamefont {Feng},
  \citenamefont {Ye},\ and\ \citenamefont {Perkins}}]{Feng2021}%
  \BibitemOpen
  \bibfield  {author} {\bibinfo {author} {\bibfnamefont {K.}~\bibnamefont
  {Feng}}, \bibinfo {author} {\bibfnamefont {M.}~\bibnamefont {Ye}}, \ and\
  \bibinfo {author} {\bibfnamefont {N.~B.}\ \bibnamefont {Perkins}},\ }\href
  {https://arxiv.org/abs/2103.14661} {\bibfield  {journal} {\bibinfo  {journal}
  {arXiv:2103.14661}\ } (\bibinfo {year} {2021})}\BibitemShut {NoStop}%
\bibitem [{\citenamefont {Zotos}\ \emph {et~al.}(1997)\citenamefont {Zotos},
  \citenamefont {Naef},\ and\ \citenamefont {Prelovsek}}]{Zotos1997}%
  \BibitemOpen
  \bibfield  {author} {\bibinfo {author} {\bibfnamefont {X.}~\bibnamefont
  {Zotos}}, \bibinfo {author} {\bibfnamefont {F.}~\bibnamefont {Naef}}, \ and\
  \bibinfo {author} {\bibfnamefont {P.}~\bibnamefont {Prelovsek}},\ }\href
  {\doibase 10.1103/PhysRevB.55.11029} {\bibfield  {journal} {\bibinfo
  {journal} {Phys. Rev. B}\ }\textbf {\bibinfo {volume} {55}},\ \bibinfo
  {pages} {11029} (\bibinfo {year} {1997})}\BibitemShut {NoStop}%
\bibitem [{\citenamefont {Hentrich}\ \emph {et~al.}(2020)\citenamefont
  {Hentrich}, \citenamefont {Hong}, \citenamefont {Gillig}, \citenamefont
  {Caglieris}, \citenamefont {\ifmmode~\check{C}\else \v{C}\fi{}ulo},
  \citenamefont {Shahrokhvand}, \citenamefont {Zeitler}, \citenamefont
  {Roslova}, \citenamefont {Isaeva}, \citenamefont {Doert}, \citenamefont
  {Janssen}, \citenamefont {Vojta}, \citenamefont {B\"uchner},\ and\
  \citenamefont {Hess}}]{Hentrich2020}%
  \BibitemOpen
  \bibfield  {author} {\bibinfo {author} {\bibfnamefont {R.}~\bibnamefont
  {Hentrich}}, \bibinfo {author} {\bibfnamefont {X.}~\bibnamefont {Hong}},
  \bibinfo {author} {\bibfnamefont {M.}~\bibnamefont {Gillig}}, \bibinfo
  {author} {\bibfnamefont {F.}~\bibnamefont {Caglieris}}, \bibinfo {author}
  {\bibfnamefont {M.}~\bibnamefont {\ifmmode~\check{C}\else \v{C}\fi{}ulo}},
  \bibinfo {author} {\bibfnamefont {M.}~\bibnamefont {Shahrokhvand}}, \bibinfo
  {author} {\bibfnamefont {U.}~\bibnamefont {Zeitler}}, \bibinfo {author}
  {\bibfnamefont {M.}~\bibnamefont {Roslova}}, \bibinfo {author} {\bibfnamefont
  {A.}~\bibnamefont {Isaeva}}, \bibinfo {author} {\bibfnamefont
  {T.}~\bibnamefont {Doert}}, \bibinfo {author} {\bibfnamefont
  {L.}~\bibnamefont {Janssen}}, \bibinfo {author} {\bibfnamefont
  {M.}~\bibnamefont {Vojta}}, \bibinfo {author} {\bibfnamefont
  {B.}~\bibnamefont {B\"uchner}}, \ and\ \bibinfo {author} {\bibfnamefont
  {C.}~\bibnamefont {Hess}},\ }\href {\doibase 10.1103/PhysRevB.102.235155}
  {\bibfield  {journal} {\bibinfo  {journal} {Phys. Rev. B}\ }\textbf {\bibinfo
  {volume} {102}},\ \bibinfo {pages} {235155} (\bibinfo {year}
  {2020})}\BibitemShut {NoStop}%
\bibitem [{\citenamefont {Baek}\ \emph {et~al.}(2020)\citenamefont {Baek},
  \citenamefont {Yeo}, \citenamefont {Do}, \citenamefont {Choi}, \citenamefont
  {Janssen}, \citenamefont {Vojta},\ and\ \citenamefont
  {B\"uchner}}]{Baek2020}%
  \BibitemOpen
  \bibfield  {author} {\bibinfo {author} {\bibfnamefont {S.-H.}\ \bibnamefont
  {Baek}}, \bibinfo {author} {\bibfnamefont {H.~W.}\ \bibnamefont {Yeo}},
  \bibinfo {author} {\bibfnamefont {S.-H.}\ \bibnamefont {Do}}, \bibinfo
  {author} {\bibfnamefont {K.-Y.}\ \bibnamefont {Choi}}, \bibinfo {author}
  {\bibfnamefont {L.}~\bibnamefont {Janssen}}, \bibinfo {author} {\bibfnamefont
  {M.}~\bibnamefont {Vojta}}, \ and\ \bibinfo {author} {\bibfnamefont
  {B.}~\bibnamefont {B\"uchner}},\ }\href {\doibase
  10.1103/PhysRevB.102.094407} {\bibfield  {journal} {\bibinfo  {journal}
  {Phys. Rev. B}\ }\textbf {\bibinfo {volume} {102}},\ \bibinfo {pages}
  {094407} (\bibinfo {year} {2020})}\BibitemShut {NoStop}%
\end{thebibliography}%
\end{document}